\documentclass[reprint,superscriptaddress,amsmath,amssymb,a4paperaps,pra,nofootinbib]{revtex4-1}
\usepackage{dcolumn,amssymb,amsmath,amsfonts,graphicx,latexsym,float,xcolor}
\usepackage{epstopdf}
\usepackage{nameref}
\usepackage{txfonts}
\usepackage{ulem}
\usepackage{easyReview}
\usepackage{tikz}
\begin{document}

\title{Intra- and interband excitations induced residue decay of the Bose polaron\\ in a one-dimensional double-well}
\author{Jie Chen}
\email {jie.chen@physnet.uni-hamburg.de}
\affiliation{Department of Physics, Center for Optical Quantum Technologies, University of Hamburg, Luruper Chaussee 149, 22761 Hamburg, Germany}
\author{Simeon I. Mistakidis}
\affiliation{ITAMP, Center for Astrophysics $|$ Harvard $\&$ Smithsonian, Cambridge, MA 02138 USA}
\author{Peter Schmelcher}
\affiliation{Department of Physics, Center for Optical Quantum Technologies, University of Hamburg, Luruper Chaussee 149, 22761 Hamburg, Germany}
\affiliation{The Hamburg Center for Ultrafast Imaging, University of Hamburg, Luruper Chaussee 149, 22761 Hamburg, Germany}
\date{\today}

\begin{abstract}
We investigate the polaronic properties of a single impurity immersed in a weakly interacting bosonic environment confined within a one-dimensional double-well potential using an exact diagonalization approach. We find that an increase of the impurity-bath coupling results in a vanishing residue, signifying the occurrence of the polaron orthogonality catastrophe. Asymptotic configurations of the systems' ground state wave function in the strongly interacting regime are obtained by means of a Schmidt decomposition, which in turn accounts for the observed orthogonality catastrophe of the polaron. We exemplify that depending on the repulsion of the Bose gas, three distinct residue behaviors appear with respect to the impurity-bath coupling. These residue regimes are characterized by two critical values of the bosonic repulsion and originate from the interplay between the intra- and the interband excitations of the impurity. Moreover, they can be clearly distinguished in the corresponding species reduced density matrices with the latter revealing a phase separation on either the one- or the two-body level. The impact of the interspecies mass-imbalance on the impurity's excitation processes is appreciated yielding an interaction shift of the residue regions. Our results explicate the interplay of intra- and interband excitation processes for the polaron generation in multiwell traps and for designing specific polaron entangled states motivating their exposure in current experiments.
\end{abstract}

\maketitle
\section{Introduction}
Ultracold atomic gases provide pristine platforms to study quantum many-body physics owing to their unprecedented controllability, e.g., in terms of the involved trapping geometries and the atomic interactions~\cite{cold_atom_rev, Feshbach_1,Feshbach_2,Feshbach_3}. Among the achieved milestones \cite{BH_exp_1,BH_exp_2,BH_exp_3}, trapping of a many-body bosonic gas in a one-dimensional (1D) double-well (DW) potential constitutes a prototype system for unravelling the emergent complex quantum dynamics \cite{DW_exp_1,DW_exp_2, DW_exp_3}. This system represents a bosonic Josephson junction (BJJ), namely the atomic analogue of the Josephson effect initially predicted for tunneling of Cooper pairs through two weakly linked superconductors~\cite{BJJ_1,BJJ_2}. Relevant investigations of the BJJ unveiled various intriguing phenomena including, for instance, Josephson oscillations \cite{BJJ_Rabi_1, BJJ_Rabi_2, BJJ_Rabi_3}, macroscopic quantum self-trapping events~\cite{DW_exp_3, BJJ_Rabi_1, BJJ_Rabi_2}, collapse and revival population sequences~\cite{BJJ_Rabi_3}, and the formation of an atomic squeezed state~\cite{BJJ_Squeeze_1, BJJ_Squeeze_2}. Moreover, the role of interparticle correlations has been examined~\cite{BJJ_Frag_1, BJJ_Frag_2}, revealing the existence of strongly correlated tunneling processes in few-body systems~\cite{BJJ_Few_1, BJJ_Few_2, BJJ_Few_3, BJJ_Few_4, BJJ_Few_5}. 
The above mechanisms are not accessible in conventional superconducting systems.

On the other hand, with the aid of sympathetic cooling as well as the experimental progresses on realizing few-body ensembles~\cite{mixture_exp_bf_1, mixture_exp_bf_2, mixture_exp_bb_1, mixture_exp_bb_2, few_exp1,few_exp2, few_exp3, few_exp4, few_exp5,few_exp6}, studies of ultracold atomic mixtures featuring an appreciably large particle number imbalance have also been put forward for both bosonic~\cite{Bose_polaron_1, Bose_polaron_2} and fermionic~\cite{Fermi_polaron_1, Fermi_polaron_2, Fermi_polaron_3} settings. 
These systems are inherently related to the so-called polaron concept~\cite{Massignan_rev,Schmidt_rev}, which is originally introduced in the context of a mobile impurity immersed in a quantum many-body environment with the former becoming dressed by the excitations of the bath~\cite{polaron_conmat_1, polaron_conmat_2, polaron_conmat_3, polaron_conmat_4}. 
So far, notable investigations of the emergent polaron properties in the ultracold atomic realm unveil, for example, their effective mass~\cite{polaron_eff_mass_1, polaron_eff_mass_2,polaron_eff_mass_3,polaron_eff_mass_Mist_Vol, Polaron_1D_exp}, the underlying excitation spectra~\cite{polaron_ext_spec_1,polaron_ext_spec_2, Mistakidis_rf, Mistakidis_pump_probe,Will}, the existence of bath mediated induced-interactions~\cite{polaron_ind_int_1,polaron_ind_int_2,polaron_ind_int_3,polaron_ind_int_4,polaron_ind_int_Mist_Vol,Mukherjee_induced_int}, as well as the orthogonality catastrophe (OC) mechanism~\cite{polaron_OC_1,polaron_OC_2,polaron_OC_3,polaron_OC_4,Mistakidis_OC_2imp}. 

Interestingly, the above investigations focus on either homogeneous \cite{polaron_eff_mass_1,polaron_eff_mass_2,polaron_eff_mass_3, polaron_ext_spec_1, polaron_ext_spec_2, polaron_OC_1, polaron_OC_4, Will} or harmonically trapped mixtures \cite{polaron_eff_mass_Mist_Vol, polaron_ind_int_1, polaron_ind_int_2, polaron_ind_int_3, polaron_ind_int_4, polaron_ind_int_Mist_Vol,Mukherjee_induced_int, polaron_OC_2, polaron_OC_3, Mistakidis_OC_2imp,Mistakidis_rf, Mistakidis_pump_probe, Polaron_1D_exp}.
Studies considering a strong  spatial inhomogeneity, e.g., by assuming that both the impurity and the bath are trapped within a DW potential are still rare \cite{polaron_few_1,polaron_few_2}. 
Indeed, current explorations have predominantly dealt with the impurity transport in lattices~\cite{Cai_transport1,Johnson_transport2} and were mainly restricted to the lowest-band approximation besides a few exceptions~\cite{Theel_higherband}. 
As such the interplay of excitation processes within the same or between different energetic bands as a result of the inhomogeneity is still far from being completely understood even in the static properties of these systems.  
Since the DW offers a toy-model of a lattice geometry where the band-structure is important, it provides a testbed for examining the interplay of related excitations and in particular their role in the polaron generation~\cite{Yin_twoband}. 
Here, the polaronic behavior should strongly depend on the involved interactions and an intriguing prospect would be to engineer specific entangled polaron states in certain interaction regimes with an additional knob provided by the interspecies mass-imbalance. 
Moreover, we should also emphasize that these hybridized systems are of further interest due to the fact that one subsystem (impurity) lies in the deep quantum regime while the other one (medium) can be potentially described semi-classically~\cite{Impurity_BH_1,Impurity_BH_2,Impurity_BH_3,Impurity_BH_4,Impurity_BH_5,Impurity_BH_6}. For instance, it has been demonstrated that following an impurity-bath interaction quench leads to chaotic signatures in the dynamics of the bath accompanied by significant coherence losses~\cite{Impurity_BH_6}.

In the present work, we investigate the polaron properties of an atomic mixture where a single impurity is embedded into a weakly interacting bosonic gas confined within a 1D DW potential. Particular focus is placed on the polaronic properties appearing in the ground state (GS) of the mixture upon considering variations of the involved interaction strengths (both the intra- and interspecies ones) as well as the interspecies mass-imbalance. To argue on the emergent polaron generation and consequent behavior, we rely on the so-called residue~\cite{polaron_OC_1,polaron_OC_2,polaron_OC_3,polaron_OC_4}, being a measure of the overlap between the dressed polaronic state and the initial non-interacting one. 
We analyze this composite system via the numerically exact diagonalization (ED) method. 
The latter allows us to take all correlations of the mixture into account as well as capture the impurity's higher-band excitation processes. 

We find that an increase of the impurity-bath repulsion results in a vanishing residue, signifying the occurrence of the polaron OC. Employing a Schmidt decomposition of the many-body wave function, we obtain two asymptotic configurations of the GS wave function for strong interspecies repulsions, which in turn account for the observed OC of the polaron. We explicate that depending on the repulsion strength of the Bose gas, three distinct residue behaviors appear with respect to the impurity-bath coupling. More specifically, with increasing interspecies repulsion the residue exhibits: i) a sharp decrease, ii) an initial decrease followed by a pronounced revival at intermediate repulsions and then vanishes for stronger impurity-bath coupling or iii) a slow monotonous decay. These residue regimes are characterized by two critical repulsion strengths of the Bose gas and originate from the interplay between the intra- and the interband excitations of the impurity.
Moreover, they are clearly visible in the corresponding species reduced density matrices, which alternatively reveal a spatial phase separation on either the one- or the two-body level. 
The impact of the mass-imbalance on the impurity's excitation processes is also discussed yielding a shift of the above-described residue regions with respect to the bosonic repulsion.

This work is organized as follows. In Sec.~\ref{Setup}, we introduce our setup and the employed tight-binding description for the bath. Sec.~\ref{Results} presents our main observation: the existence of three  interaction regions characterized by distinct polaronic residue behaviors. We describe in detail each residue regime in Sec.~\ref{Type-I_residue_decay}, Sec.~\ref{Type-II_residue_decay} and Sec.~\ref{Type-III_residue_decay}, respectively. Afterwards, a spectral analysis is performed in Sec.~\ref{two_bosonic_repulsion} unveiling the origin of the underlying critical bosonic repulsions separating the aforementioned residue regimes. An outlook is provided in Sec.~\ref{Conclusions} containing some future perspectives of our findings. 
In Appendix~\ref{Appendix_parity} we discuss the impact of the 
parity conservation in the GS wave function on the shape of the bosonic density configuration.
Appendix~\ref{Appendix_A} explicates the form of the asymptotic wave function in terms of the Wannier basis representation, while Appendix~\ref{Appendix_B} elaborates on the ingredients of the species mean-field (SMF) approach.

\section{Impurity setting and tight-binding description} \label{Setup}
We consider a highly particle imbalanced binary atomic mixture consisting of a single impurity $N_{I} = 1$ being immersed in a bosonic gas containing $N_{B} = 100$ bosons. As such for finite impurity-bath couplings the impurity is dressed by the excitations of the bosonic gas forming a Bose polaron~\cite{Polaron_1D_exp}, a behavior that has been exemplified in 1D by the emulation of relevant spectroscopy schemes~\cite{Mistakidis_rf,Mistakidis_OC_2imp,Mistakidis_pump_probe} as well as by constructing related effective polaron Hamiltonians~\cite{polaron_eff_mass_Mist_Vol}. The underlying many-body Hamiltonian is given by $\hat{H} = \hat{H}_{I} + \hat{H}_{B} +  \hat{H}_{IB}$, where 
\begin{align}
\hat{H}_{I} &=\int dx~\hat{\psi}^{\dagger}_{I}(x) \textit{h}_{I}(x) \hat{\psi}_{I}(x), \nonumber\\
\hat{H}_{B} &=\int dx~\hat{\psi}^{\dagger}_{B}(x) \textit{h}_{B}(x) \hat{\psi}_{B}(x) \nonumber\\
&+\frac{g_{BB}}{2}\int dx~\hat{\psi}^{\dagger}_{B}(x)\hat{\psi}^{\dagger}_{B}(x)\hat{\psi}_{B}(x)\hat{\psi}_{B}(x), \nonumber \\
\hat{H}_{IB} &= {g_{IB}}\int dx~\hat{\psi}^{\dagger}_{I}(x) \hat{\psi}^{\dagger}_{B}(x) \hat{\psi}_{B}(x) \hat{\psi}_{I}(x). \label{Hamiltonian_IB}
\end{align}
In the above expressions, $\textit{h}_{\sigma}(x) = -\frac{\hbar^{2}}{2 m_{\sigma}}\frac{\partial^{2}}{\partial x^{2}}+ V_{DW}(x)$ is the single-particle Hamiltonian for the $\sigma = I(B)$ species being confined within a 1D symmetric DW potential $V_{DW}(x) = a_{\sigma} (x^{2} - b_{\sigma}^{2})^{2}$. The parameters $a_{\sigma}$ and $b_{\sigma}$ control the central barrier height as well as the relative distance between the two wells, respectively. $ \hat{\psi}_{\sigma}^{\dagger}(x)$ [$\hat{\psi}_{\sigma}(x)$] is the field operator that creates (annihilates) a $\sigma$-species particle of mass $m_{\sigma}$ at position $x$~\cite{Pitaevskii_book}. Moreover, we assume that both the intra- (Bose-Bose) and the interspecies (impurity-bath) interactions are short-range i.e. of contact type characterized by the strengths $g_{BB}$ and $g_{IB}$ respectively. This is an adequate approximation within the ultracold temperature regime where $s$-wave interaction processes are the dominant ones~\cite{Feshbach_1}. In the following, we rescale the Hamiltonian of the mixture $\hat{H}$ in harmonic units, with the energy and length being expressed in terms of $\eta = \hbar  \omega$ and $\xi = \sqrt{\hbar /m_{B} \omega}$, respectively. 

\begin{figure}
	\centering
	\includegraphics[width=0.5\textwidth]{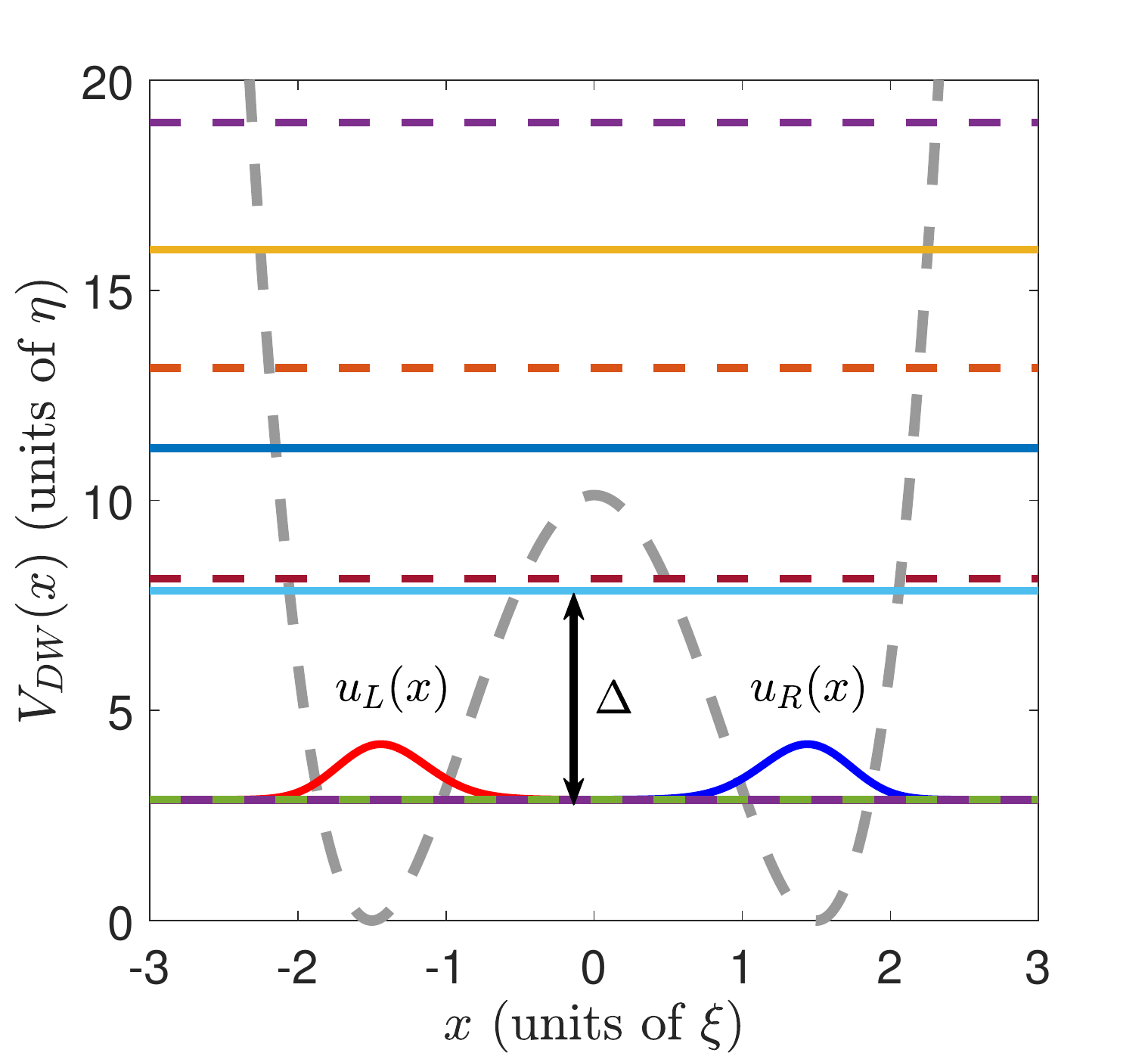}\hfill
	\caption{(Color online) Schematic representation of the DW (grey dashed line) characterized by $a_{DW} = 2.0$, $b_{DW} = 1.5$ and its energetically lowest single-particle spectrum. The horizontal lines denote the lowest eight energy levels, in which the solid (dashed) lines correspond to the even (odd) parity states of the DW. The energy difference (gap) between the lowest two energetic bands is denoted by $\Delta$. The red and blue solid lines represent the Wannier states $u_{L}(x)$ and $u_{R}(x)$ in the lowest band of the left and right well, respectively.} 
	\label{DW_potential}
\end{figure}

Throughout this work, we assume that the atoms of both species are trapped in the same DW geometry, i.e., $a_I = a_B = a_{DW}$ and $b_I = b_B = b_{DW}$. In practice, we set $a_{DW} = 2.0$ and $b_{DW}=1.5$ such that the low-lying single-particle energy levels form a band-like doublet structure. The spatial geometry of $V_{DW}(x)$ is depicted in Fig.~\ref{DW_potential} (see grey dashed line) together with the lowest eight single-particle energy levels (see horizontal lines). Moreover, we focus on repulsive interactions for both species and assume that the bosonic bath is weakly interacting, i.e., $0< g_{BB} \ll 1$. We explore how the GS polaronic nature is affected by considering variations of both interaction strengths as well as the mass-imbalance $\beta = m_{I}/ m_{B}$ between the two species. Let us note that such a 1D mixture is experimentally accessible by imposing a strong transverse and a weak longitudinal confinement to a binary e.g., Bose-Fermi mixture with two different kinds of atoms \cite{mixture_exp_bf_1, mixture_exp_bf_2} or a Bose-Bose mixture that is made of atoms residing in two different hyperfine states \cite{mixture_exp_bb_1, mixture_exp_bb_2}. The DW potential can also be readily constructed by imposing a 1D optical lattice on top of a harmonic trap, constructing a superlattice whose shells correspond to double-wells \cite{DW_exp_3, BJJ_2}. Moreover, the contact interaction strengths $g_{IB}$ and $g_{BB}$ can be controlled experimentally by tuning the $s$-wave scattering lengths via Feshbach or confinement-induced resonances \cite{Feshbach_1,Feshbach_2,Feshbach_3}.

Given that the bosons are weakly interacting and are confined within a tight DW with a large energy gap $\Delta$ between the first and the second band~\footnote{For the employed DW parameters, $a_{DW} = 2.0$ and $b_{DW} = 1.5$, the energy gap $\Delta$ between the first and the second band is approximately $10^3$ times larger than the width of the first band.} [cf. Fig.~\ref{DW_potential}], we further adopt the two-mode approximation for the bosonic bath
\begin{equation}
\hat{\psi}_{B}(x) = u_{L}(x) \hat{b}_{L} + u_{R}(x) \hat{b}_{R}, \label{2_mode_psi}
\end{equation}
with $u_{L,R}(x)$ being the Wannier states~\cite{BJJ_2} localized in the left and the right well, respectively [cf. Fig.~\ref{DW_potential}]. They are chosen to be real valued due to the preserved time-reversal symmetry of the mixture \cite{quantum_mechanics_Landau}. This ansatz leads to the low-energy effective Hamiltonian for the bosonic species
\begin{equation}
\hat{H}_{B} =  -J_{B} (\hat{b}^{\dagger}_{L}\hat{b}_{R} + \hat{b}^{\dagger}_{R} \hat{b}_{L}) + \frac{U_{B}}{2} \sum_{i = L,R} \hat{b}^{\dagger}_{i}\hat{b}^{\dagger}_{i}\hat{b}_{i}\hat{b}_{i}, \label{BH_model}
\end{equation}
corresponding to the two-site Bose-Hubbard (BH) model \cite{2_site_BH_1, 2_site_BH_2} with 
\begin{align}
	J_{B} &= \int dx ~ u_{L}(x)\textit{h}_{B}(x)u_{R}(x), \nonumber \\
	U_{B} &= g_{BB}\int dx ~ |u_{i}(x)|^{4} ~~~~~ (i = L,R),
\end{align}
representing the hopping amplitude and the on-site repulsion energy, respectively.

Hereafter, we designate the GS of our mixture for $g_{IB}>0$ ($g_{IB} = 0$) by $|\Psi\rangle$ ($|\Psi_{0}\rangle$), and denote the eigenstates of the single-species Hamiltonian $\hat{H}_{I}$ and $\hat{H}_{B}$ as $\{|\phi_{i}^{I}\rangle\}$ and $\{|\phi_{i}^{B}\rangle\}$, respectively. Moreover, we refer to the eigenstates of the bosonic single-particle Hamiltonian $h_{B}$ as $\{|\varphi^{B}_{i}\rangle \}$. In this way, the above introduced Wannier states $|u_{L,R} \rangle$ are expressed with respect to $|\varphi^{B}_{0}\rangle$ and $|\varphi^{B}_{1}\rangle$ as $|u_{L,R}\rangle = \frac{1}{\sqrt{2}} [|\varphi^{B}_{0}\rangle \pm |\varphi^{B}_{1}\rangle]$. The preserved parity symmetry of the DW potential results in each eigenstate of $\hat{H}$ or $\hat{H}_{\sigma}$ to possess a definitive parity. Specifically, the GS $|\Psi\rangle$ ($|\Psi_{0}\rangle$) is even parity, while $|\phi_{i}^{\sigma}\rangle$ for $i = 0,2,4,\dots$ ($i = 1,3,5,\dots$) possesses an even (odd) parity.

\section{Bose polaron residue response and its decay} \label{Results}
In the following, we investigate the GS polaronic properties of the mixture. In particular, we focus on how variations of both the intra- and the interspecies interactions as well as the mass-imbalance impact the emergent polaronic nature. To this end, we examine as a representative measure the so-called polaronic residue, quantified by the overlap between the dressed polaronic state and the initial non-interacting one \cite{polaron_OC_1,polaron_OC_2,polaron_OC_3,polaron_OC_4}. As we shall argue below, the increase of $g_{IB}$ always leads to a vanishing residue, signifying the occurrence of the OC of the polaron. Interestingly, we shall also exemplify that there exist two critical values of the bosonic repulsion, referred to as $g_{BB}^{cI}$ and $g_{BB}^{cII}$ (depending also on the mass-imbalance $\beta$) according to which three distinct residue response regions are encountered for increasing $g_{IB}$. We perform a detailed investigation demonstrating the properties of these regions and explicating the origin of the aforementioned critical bosonic repulsion strengths.

\subsection{Overview of the polaronic residue} \label{structure_factor}
\begin{figure*}
	\centering
	\includegraphics[width=1.0\textwidth]{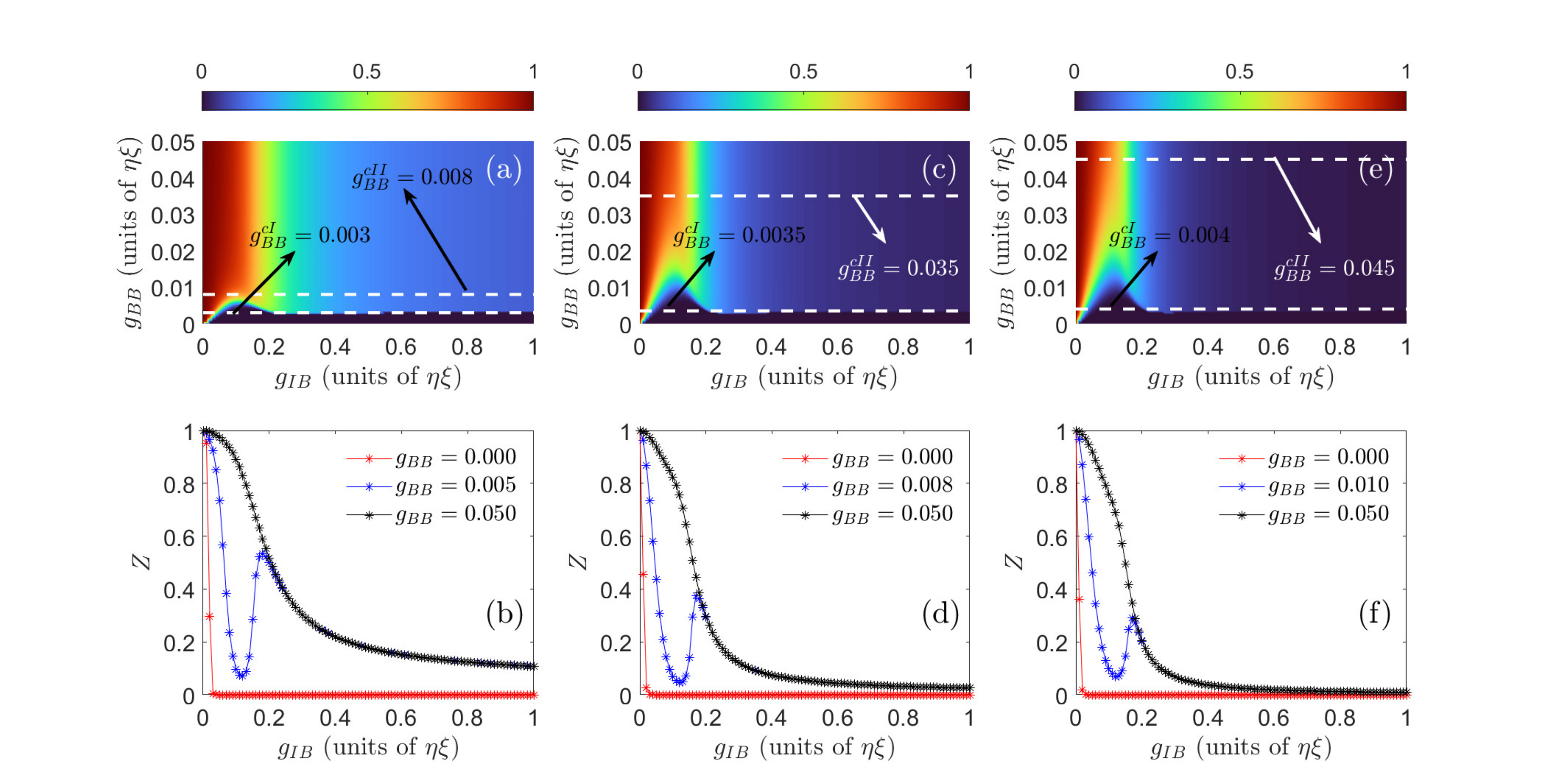}\hfill
	\caption{(Color online) Polaronic residue $Z$ with respect to variations of the involved interaction strengths and particular mass ratios. The latter refer to (upper panels) (a) $\beta = 0.5$, (c) $\beta = 1.0$ and (e) $\beta = 2.0$. Three different residue regimes in terms of $g_{BB}$ are identified and they are separated by the white dashed horizontal lines, which denote the critical bosonic repulsion strengths corresponding to (a) $g_{BB}^{cI} = 0.003$, $g_{BB}^{cII} = 0.008$, (c) $g_{BB}^{cI} = 0.0035$, $g_{BB}^{cII} = 0.035$ and (e) $g_{BB}^{cI} = 0.004$, $g_{BB}^{cII} = 0.045$. Lower panels: same with the upper panels but for specific bosonic repulsions (see legends). The residue decay designating the OC of the Bose polaron for strong repulsive $g_{IB}$ is evident. }
	\label{residual_Z}
\end{figure*}

The polaronic residue is given by $Z = |\langle \Psi_{0}|\Psi\rangle|^{2}$, which measures the overlap between the dressed polaronic state and the initial non-interacting one. Here, $|\Psi_{0}\rangle = |\phi^{I}_{0} \rangle |\phi^{B}_{0} \rangle$ is the GS of the mixture for fixed $g_{BB}$ and $\beta$ and vanishing impurity-medium coupling ($g_{IB} = 0$), while $|\Psi\rangle$ denotes the GS for $g_{IB} > 0$. In Fig.~\ref{residual_Z}, we present the residue $Z$ as a function of both the Bose-Bose and the impurity-bath interaction strengths for various mass ratios, namely $\beta=0.5$ [Fig.~\ref{residual_Z} (a,b)], $\beta=1.0$ [Fig.~\ref{residual_Z} (c,d)] and $\beta=2.0$ [Fig.~\ref{residual_Z} (e,f)]. As it can be seen, for fixed $\beta$ and $g_{BB}$, an increase of $g_{IB}$ leads eventually to the suppression of the polaronic residue, which in turn implies the decay of the polaronic nature. In particular, the residue becomes negligible for $g_{IB} \rightarrow 1$, signifying that the OC of the polaron takes place in this strongly interacting regime \footnote{Note that the mechanism of the polaron OC, discussed herein, is different from the original Anderson OC \cite{polaron_conmat_4}. The latter is strictly defined in the thermodynamic limit while our finite atom number mixture is also spatially inhomogeneous due to the DW. Instead, we rely on the fact that for a GS with vanishing polaronic nature its residue becomes $Z  = 0$, a behavior that it is here partially caused by the trap-induced spatial inhomogeneity.  
Albeit this differentiation, as we shall see below, the polaronic residue in our system exhibits a similar behavior to the one obtained in the Anderson's original argument. Namely, for an increasing bath particle number $N_{B}\rightarrow \infty$ the residue $Z \rightarrow 0$.}. Moreover, for all different mass ratios, there exist two critical bosonic repulsion strengths $g_{BB}^{cI}$ and $g_{BB}^{cII}$ [see the white dashed horizontal lines in Fig.~\ref{residual_Z}]. Depending on $g_{BB}$ being below or above each of these critical values, three distinct residue regions are found for increasing $g_{IB}$. For instance, in the equal-mass case we have $g_{BB}^{cI} = 0.0035$ and $g_{BB}^{cII} = 0.035$ [Fig.~\ref{residual_Z} (c)]. For $g_{BB} = 0$ and increasing $g_{IB}$, the residue, starting from $Z =1$, sharply drops to $Z \approx 0$ for $g_{IB} = 0.003$ and remains almost constant for further increasing $g_{IB}$ [cf. Fig.~\ref{residual_Z} (d) red line]. This behavior essentially reveals that, for $g_{BB} < g_{BB}^{cI}$, the OC of the polaron takes place even for an extremely weak impurity-bath repulsion. In contrast, once we ramp up the bosonic repulsion e.g., to $g_{BB} = 0.008$, the residue initially decreases to $Z = 0.05$ and then exhibits a pronounced peak for larger $g_{IB}$ [cf. Fig.~\ref{residual_Z} (d) blue line],  which signifies the revival of the polaron for intermediate impurity-bath interactions and within the region $g_{BB} \in [g_{BB}^{cI}, g_{BB}^{cII}]$. Turning to $g_{BB} > g_{BB}^{cII}$, a relatively slower monotonous decrease of $Z$ with respect to $g_{IB}$ is observed, suggesting that the mixture enters a regime with a slow polaron decay [see black line in Fig.~\ref{residual_Z} (d) for the case $g_{BB} = 0.05$].

Interestingly, while $g_{BB}^{cI}$ appears to be inert with respect to variations of the mass ratio, $g_{BB}^{cII}$ depends strongly on $\beta$, e.g., $g_{BB}^{cII} = 0.008$ for $\beta = 0.5$ while $g_{BB}^{cII} = 0.035$ for $\beta = 1.0$ and $g_{BB}^{cII} = 0.045$ for $\beta = 2.0$. Based on the above observations, we will refer to the aforementioned three distinct regions as type-I ($g_{BB} \in [0,g_{BB}^{cI}]$), type-II ($g_{BB} \in [g_{BB}^{cI},g_{BB}^{cII}]$) and type-III polaron decay ($g_{BB} > g_{BB}^{cII}$), respectively. Moreover, in order to analyze the properties and origin of these regions we will solely focus on the equal-mass case [Sec.~\ref{Type-I_residue_decay} to Sec.~\ref{Type-III_residue_decay}], while the influence of the mass-imbalance will be discussed later on in Sec.~\ref{two_bosonic_repulsion}.

\subsection{Type-I polaron decay} \label{Type-I_residue_decay}
\begin{figure*}
	\centering
	\includegraphics[width=1.0\textwidth]{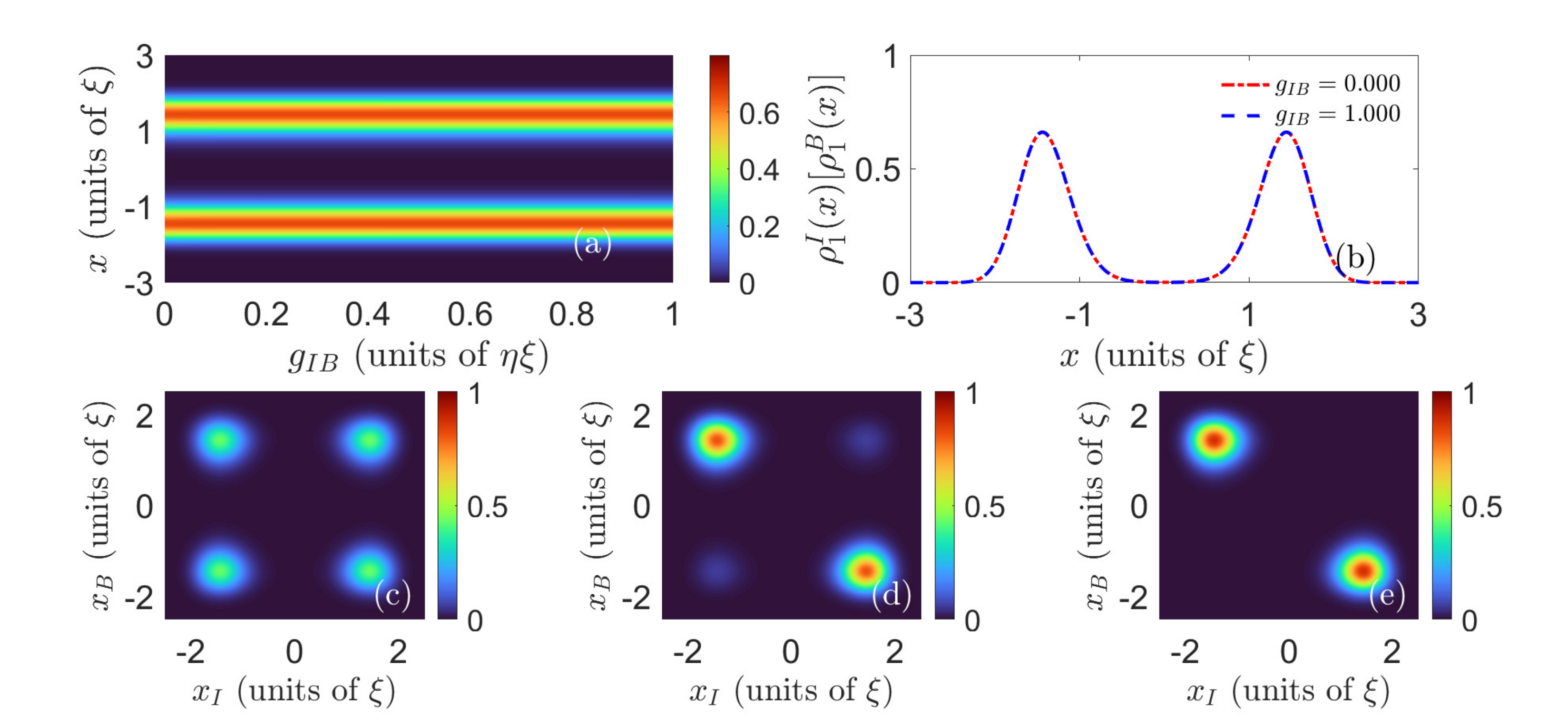}\hfill
	\caption{(Color online) Spatial profiles of the impurity's reduced densities for the type-I residue decay. (a) The single-particle density $\rho^{I}_{1}(x)$ of the impurity as a function of $g_{IB}$ with $g_{BB} = 0$ showing an equal population of the two wells. The bosonic gas density $\rho^{B}_{1}(x)$ exhibits exactly the same shape (not shown). (b) Profiles of $\rho^{I}_{1}(x)$ ($\rho^{B}_{1}(x)$) for $g_{BB} = 0$ and for either $g_{IB} = 0$ (red dash-dotted line) or $g_{IB} = 1.0$ (blue dashed line). (c)-(e) The impurity-bath two-body density distributions $\rho^{IB}_{2}(x_{I}, x_{B})$ in the case of $g_{BB} = 0$ and (c) $g_{IB} = 0.0$, (d) $g_{IB} = 0.003$ and (e) $g_{IB} = 1.0$. 
	A two-body phase-separation process occurs for increasing $g_{IB}$.}
	\label{dmats_type_I}
\end{figure*}

To understand the polaron decay in the interaction region $g_{BB} < g_{BB}^{cI}$, we first investigate the $\sigma$-species single-particle density $\rho^{\sigma}_{1}(x) = \langle\Psi|\hat{\psi}^{\dagger}_{\sigma}(x) \hat{\psi}_{\sigma}(x) |\Psi\rangle / N_{\sigma}$
for $\sigma = I(B)$. It provides the probability of finding a $\sigma$-species particle at position $x$ \cite{dmat_1, dmat_2} and is experimentally accessible via an average over a sample of single-shot measurements~\cite{cold_atom_rev,Disipative_motion_imp}. 
Figures~\ref{dmats_type_I} (a), (b) depict the spatial distributions of $\rho^{\sigma}_{1}(x) $ for fixed $g_{BB} = 0$ and varying $g_{IB}$. Due to the DW confinement, the $\rho^{\sigma}_{1}(x)$ configurations in the case of $g_{IB} = 0$ exhibit a two-hump structure, which reflects the spatial inhomogeneity of the mixture [cf. Fig.~\ref{dmats_type_I} (b) red dash-dotted line]. Interestingly, for increasing $g_{IB}$ the profiles of $\rho^{\sigma}_{1}(x) $ for both species remain un-altered [cf. Fig.~\ref{dmats_type_I} (a)]. We note that the behavior of $\rho^{B}_{1}(x)$ can be understood via the adopted two-mode approximation [cf. Eq.~\eqref{2_mode_psi}]. Due to the negligible spatial overlap between the two Wannier states, i.e.  $ u_{L}(x) u_{R}(x) \approx 0$,  it holds that $\rho^{B}_{1}(x) \approx  \sum_{i = L,R}  u_{i}(x) u_{i}(x) \langle\Psi| \hat{b}^{\dagger}_{i}\hat{b}_{i} |\Psi\rangle /N_{B}$. 
Together with the property $\hat{b}^{\dagger}_{L}\hat{b}_{L}  = \hat{b}^{\dagger}_{R}\hat{b}_{R}  = N_{B}/2$,  originating from the parity symmetry (see Appendix \ref{Appendix_parity} and the discussions below), $\rho^{B}_{1}(x)$ is thus invariant against variations of both $g_{IB}$ and $g_{BB}$ \footnote{This result relies on the fact that the two Wannier states have negligible spatial overlap $ u_{L}(x) u_{R}(x) \approx 0$, which is only valid for a DW with an adequately large barrier. For a shallow DW, the bosonic density indeed depends on both $g_{IB}$ and $g_{BB}$ (results are not shown).}. In contrast, the interaction dependence of the impurity's density is not explained straightforward. As we shall discuss below, this behavior can also be understood if the impurity is predominately restricted to the lowest band of the DW.

Despite the fact that the single-particle densities of the two species remain miscible, with increasing $g_{IB}$, we note that a phase separation occurs on the two-body level. This becomes evident by inspecting the two-body interspecies density distribution, i.e. $\rho^{IB}_{2}(x_{I}, x_{B}) = \langle\Psi|\hat{\psi}^{\dagger}_{I}(x_{I}) \hat{\psi}_{I}(x_{I}) \hat{\psi}^{\dagger}_{B}(x_{B}) \hat{\psi}_{B}(x_{B}) |\Psi\rangle/(N_{I}N_{B})$, which refers to the probability of detecting the impurity at position $x_I$ while one boson is at position $x_B$ \cite{dmat_1, dmat_2}. In this way, it naturally incorporates impurity-bath spatial correlations. For $g_{IB}= 0$, the two species are fully decoupled, namely both the impurity and the bosons can freely tunnel between the two wells. As a result, $\rho^{IB}_{2}(x_{I}, x_{B})$ exhibits four dominant peaks within the spatial regions ($x_{I}, x_{B}) = (\pm1.5, \pm1.5$), i.e., around the minima of the DW [cf. Figs.~\ref{DW_potential} and \ref{dmats_type_I} (c)]. With increasing $g_{IB}$, the emergent repulsion renders such a two-body density distribution energetically unfavorable.  Consequently, the impurity and the bosons tend to become anti-bunched, leading to an increase (decrease) of $\rho^{IB}_{2}(x_{I}, x_{B})$ values in the vicinity of the off-diagonal (diagonal) [cf. Fig.~\ref{dmats_type_I} (d)]. Moreover, in the case of $g_{IB} = 1.0$, the two species are fully anti-bunched, leaving a probability distribution solely along the off-diagonal [see Fig.~\ref{dmats_type_I} (e)]. As a matter of fact, the mixture in turn resides in a superposition of two equally-weighted configurations and thus forms a Schrödinger-cat state~\cite{Impurity_BH_2}, manifesting the dominant role of impurity-bath entanglement (see also below). 
Specifically, each of the two configurations represents a scenario where the impurity lies in one well while all the bosons are located in the other well. In this way, a two-body interspecies phase separation takes place in space (see also the discussion below), a phenomenon that has already been observed in fermionic mixtures~\cite{Erdmann_2b_separation}. Comparing the residue behavior  [Fig.~\ref{residual_Z} (d), red solid line] to the respective two-body densities [Fig.~\ref{dmats_type_I} (c)-(e)], it becomes clear that the OC of the polaron is directly related to the phase separation between the two species. This has been indeed confirmed independently by recent studies on the polaron dynamics, where this phase separation leads to a dissipative motion of the impurity accompanied by an energy transfer to the bath species, and hence diminishes its polaronic nature~\cite{polaron_OC_3,Disipative_motion_imp}. 

Having investigated the impact of $g_{IB}$ on the $\sigma$-species density distributions, we now analyze the structure of the many-body wave function in order to gain deeper insights into the above-described polaron decay. To this end, we first employ a Schmidt decomposition and express the GS wave function of the mixture in the form \cite{Schmidt}
\begin{equation}
	|\Psi \rangle = \sum_{i=1}^{\infty} \sqrt{\lambda_{i}} ~|\psi_{i}^{I}\rangle  |\psi_{i}^{B}\rangle, \label{psi}
\end{equation}
where $ \lambda_{i} $, being real positive values, are the Schmidt numbers with $\lambda_1 > \lambda_2 > \cdots$. They obey the constraint $\sum_{i} \lambda_i = 1$ which originates from the normalization of the wave function. $|\psi_{i}^{\sigma}\rangle $ denotes the $i$th Schmidt orbital for the $\sigma$-species and can always be expanded as a linear superposition of the $\hat{H}_{\sigma}$ eigenstates 
\begin{equation}
|\psi_{i}^{\sigma}\rangle = \sum_{k=0}^{D_{\sigma}-1} C^{\sigma}_{i,k} |\phi_{k}^{\sigma}\rangle. \label{SO_expansion}
\end{equation} 
Here, $D_{\sigma}$ is the dimension of the Hilbert space corresponding to the $\sigma$-species Hamiltonian $\hat{H}_{\sigma}$. Regarding the bosonic species, $D_{B}$ is finite due to the employed two-mode approximation and we have $D_{B} = N_B+1 = 101$. For the impurity, we truncate $D_{I}$ to some finite large values depending on the interaction strengths in order to guarantee the convergence of our results.  
It is important to note that the Schmidt numbers $\{\lambda_{i}\}$ directly reveal the presence of interspecies entanglement~\cite{Schmidt}. In particular, if $\lambda_{1} = 1$ while the remaining $\lambda_{i \neq 1} = 0$, the mixture is non-entangled with the corresponding wave function being of a simple product form, i.e.~$|\Psi \rangle = |\psi^{I} \rangle |\psi^{B} \rangle$ (for simplicity, we have neglected the subscript of each Schmidt orbital for this product form).  
As we shall discuss later, the mixture in this special limit is fully captured by the SMF description, where the mutual impact of the species is merely an effective potential \cite{polaron_ind_int_3}.

\begin{figure}
  \centering
  \includegraphics[width=0.5\textwidth]{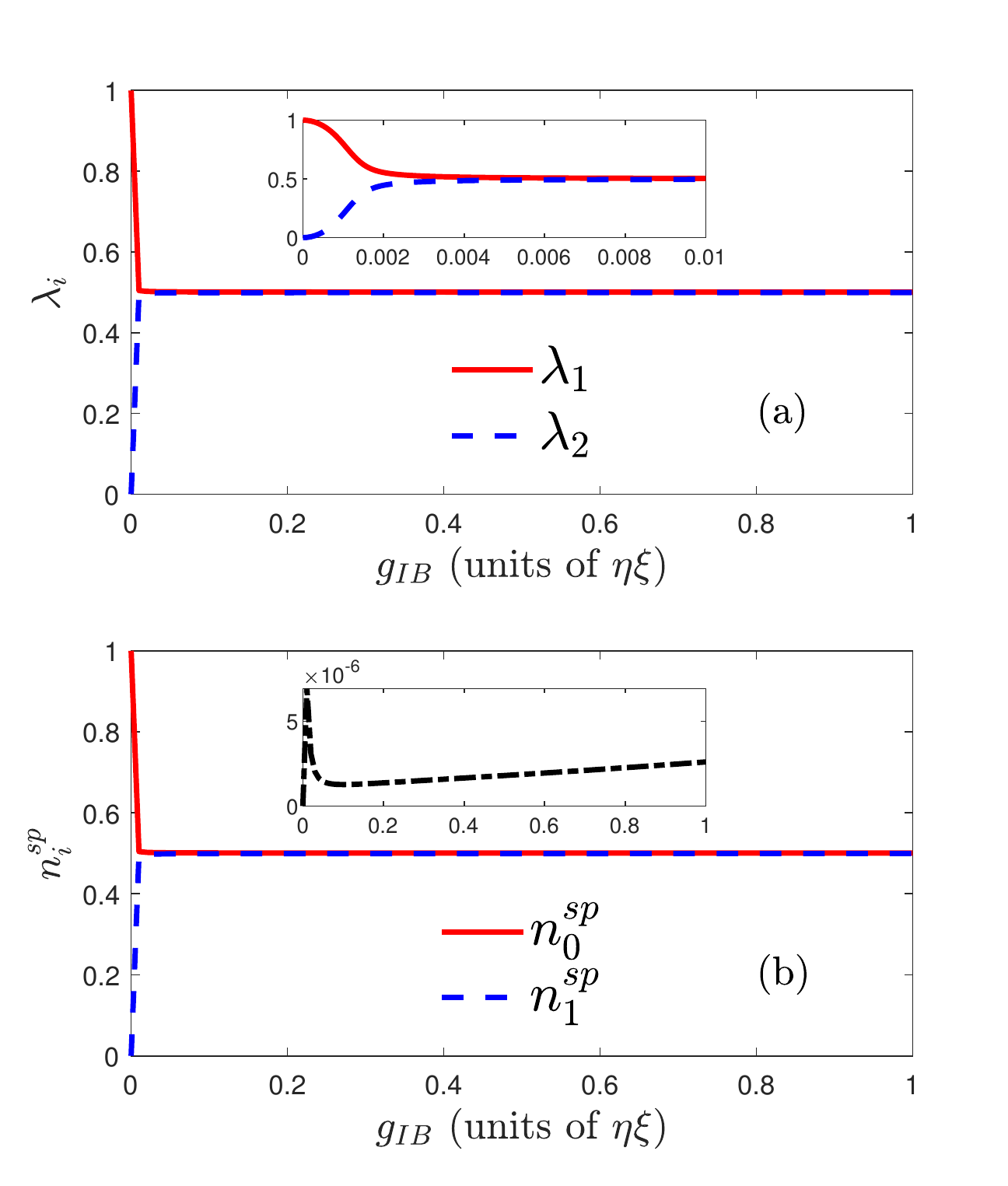}\hfill
  \caption{(Color online) Schmidt numbers and SP occupations of the impurity quantifying interspecies entanglement and the excitations of the impurity. (a) The first two Schmidt numbers for $g_{BB} = 0.0$ and varying $g_{IB}$. The red solid (blue dashed) line denotes the value of $\lambda_{1}$ ($\lambda_{2}$). 
  The inset provides a magnification within the region $0<g_{IB}<0.01$. (b) The SP occupation $n_{i}^{sp}$ of the impurity when considering $g_{BB}=0$ as a function of $g_{IB}$. The red solid (blue dashed) line refers to $n_{0}^{sp}$ ($n_{1}^{sp}$). The inset depicts the SP occupations for $\sum _{i>1}n_{i}^{sp}$.}
\label{EE_band_Impurity_type_I}
\end{figure}

Figure~\ref{EE_band_Impurity_type_I} (a) showcases the first two Schmidt numbers with respect to $g_{IB}$ for the case of $g_{BB} = 0$. Note, however, that all other Schmidt numbers are explicitly excluded due to their negligible values ($\lambda_{i} < 10^{-5}$ for $i>2$). For $g_{IB} = 0$, the GS is simply $|\Psi_{0} \rangle = |\phi_{0}^{I}\rangle |\phi_{0}^{B}\rangle$, and hence the two Schmidt numbers are $\lambda_{1} = 1$ and $\lambda_{2} = 0$. Upon switching on the impurity-bath repulsion, we observe that $\lambda_{1}$ ($\lambda_{2}$) quickly decreases (increases) to the value $\lambda_{1} = 0.52$ ($\lambda_{2} = 0.48$) for $g_{IB} = 0.003$. Afterwards, it exhibits an extremely slow decrease (increase) towards $\lambda_{1} = \lambda_{2} = 0.5$ until $g_{IB} = 0.8$ and retains this value for $g_{IB}>0.8$. In this regard, we can deduce the asymptotic form of the GS wave function in the region $g_{BB} < g_{BB}^{cI}$ and for a strong impurity-bath repulsion (namely $g_{IB} > 0.8$)
\begin{equation}
	|\Psi_{T-I} \rangle = \frac{1}{\sqrt{2}} \left[|\psi_{1}^{I}\rangle |\psi_{1}^{B}\rangle +|\psi_{2}^{I}\rangle  |\psi_{2}^{B}\rangle \right], \label{psi_type_I}
\end{equation}
with the prefactor $1/\sqrt{2}$ stemming from the fact that $\lambda_{1} = \lambda_{2} = 1/2$ for large $g_{IB}$ [see also Fig.~\ref{EE_band_Impurity_type_I} (a)]. Further inspecting the first two Schmidt orbitals of the impurity, we find that 
\begin{equation}
|\psi_{1}^{I}\rangle \approx |\phi_{0}^{I}\rangle, ~~~ |\psi_{2}^{I}\rangle \approx |\phi_{1}^{I}\rangle. \label{NO_I_type_I}
\end{equation}
Together with Eq.~\eqref{psi_type_I}, this immediately yields the single-particle (SP) occupations for the impurity as $\hat{n}_{0}^{sp} = \hat{n}_{1}^{sp} \approx 1/2$ and $\hat{n}_{i>1}^{sp} \approx 0$, reflecting the fact that the impurity is predominately occupying the lowest band of the DW [cf. Fig.~\ref{EE_band_Impurity_type_I} (b) and the discussions below]. Here, $\hat{n}_{i}^{sp} = (\hat{a}_{i}^{I})^{\dagger}\hat{a}_{i}^{I}$ denotes the occupation number of the $i$th single-particle state $|\phi^{I}_{i}\rangle$ of the impurity.

\begin{figure}
	\centering
	\includegraphics[width=0.5\textwidth]{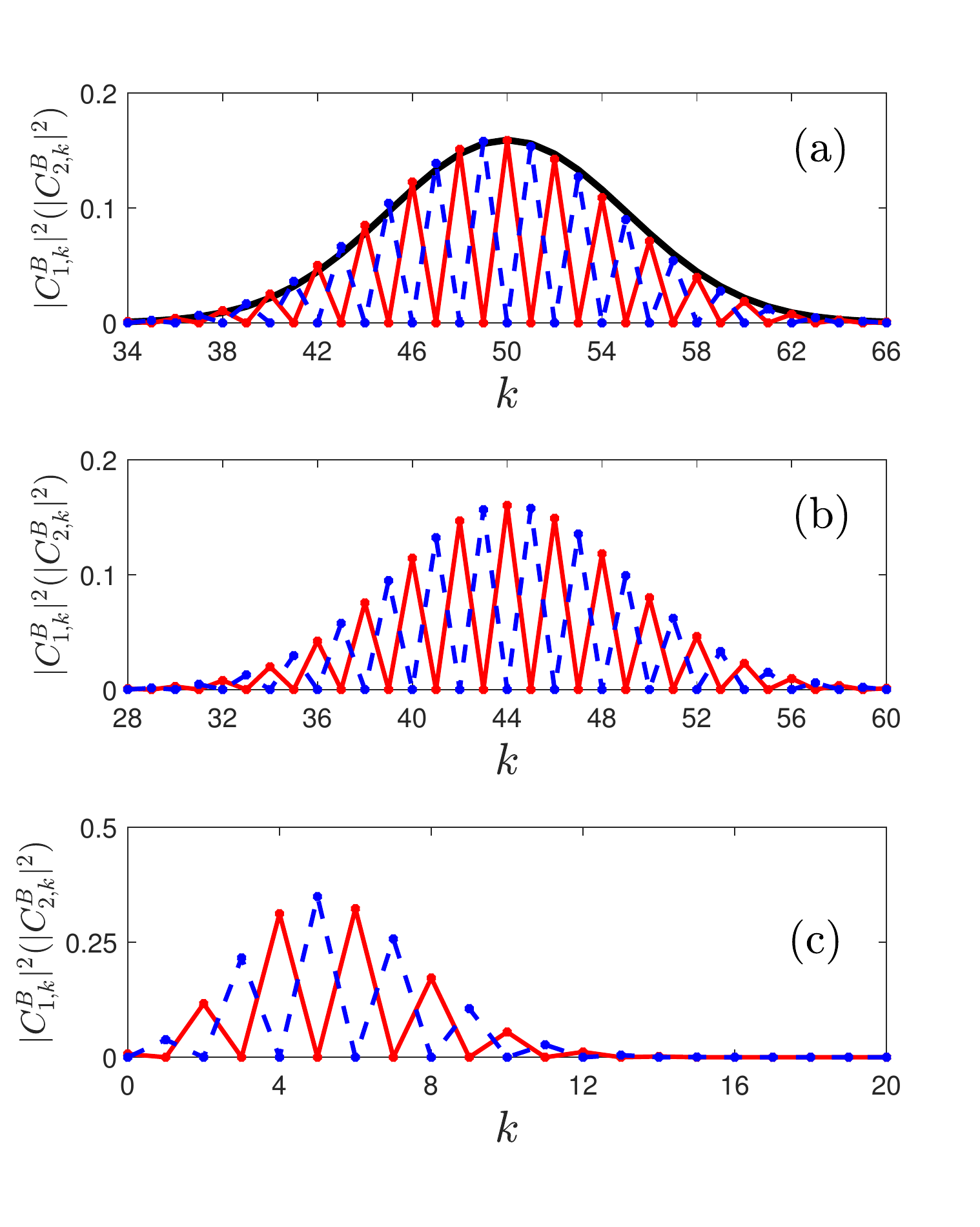}\hfill
	\caption{(Color online) Expansion coefficients for the first two bosonic Schmidt orbitals for $g_{BB} = 0.0$ and (a) $g_{IB} = 1.0$, (b) $g_{IB} = 0.05$ and (c) $g_{IB} = 0.003$. In all cases, the red solid and the blue dashed line denote $|C^{B}_{1,k}|^{2}$ and $|C^{B}_{2,k}|^{2}$, respectively. The black solid line in (a) depicts the expansion coefficients $|\tilde{C}^{B}_{1,k}|^{2}$ and  $|\tilde{C}^{B}_{2,k}|^{2}$ obtained from  Eq.~\eqref{expansion_NO_B_LR} given in Appendix \ref{Appendix_A}.}
	\label{Sorb_BH_NS_basis_type_I}
\end{figure}

On the other hand, the corresponding Schmidt orbitals for the bosonic species can be written as
\begin{equation}
|\psi_{1}^{B}\rangle = \sum_{k=0,2,4,\dots} C^{B}_{1,k} |\phi_{k}^{B}\rangle,~~~~|\psi_{2}^{B}\rangle = \sum_{k=1,3,5,\dots} C^{B}_{2,k} |\phi_{k}^{B}\rangle, \label{NO_B_type_I}
\end{equation} 
with $C^{B}_{1,k}$ and $C^{B}_{2,k}$ being the corresponding expansion coefficients and $|\phi_{k}^{B}\rangle$ denoting the $k$th eigenstate of $\hat{H}_{B}$ for the case $g_{BB} = 0$ [cf. Eq.~\eqref{SO_expansion}] . As it can be seen in Fig.~\ref{Sorb_BH_NS_basis_type_I} (a), for $g_{IB} = 1.0$, $C^{B}_{1,k}$ ($C^{B}_{2,k}$) obtained from the ED method vanishes exactly for the odd (even) parity eigenstates $|\phi_{k}^{B}\rangle$. This is the reason for the index notation $k = 0,2,4,\dots$ ($k = 1,3,5,\dots$) introduced in Eq.~\eqref{NO_B_type_I}. Physically, this observation is a direct consequence of the preserved parity symmetry of the wave function of the mixture. Indeed, since the GS wave function $|\Psi\rangle$ is of even parity, each product state $|\psi_{i}^{I}\rangle |\psi_{i}^{B}\rangle$ in Eq.~\eqref{psi_type_I} then needs to has an even parity as well. As a result, the bosonic Schmidt orbital $|\psi_{i}^{B}\rangle$ shares the same parity symmetry with the impurity. According to Eq.~\eqref{NO_I_type_I}, the Schmidt orbital $|\psi_{1}^{B}\rangle$ ($|\psi_{2}^{B}\rangle$) for the bosonic species hence 
has an even (odd) parity. Moreover, the even parity of the GS wave function $|\Psi \rangle$ ensures that the distribution of the bosons between the two wells obeys  $\hat{b}^{\dagger}_{L}\hat{b}_{L}  = \hat{b}^{\dagger}_{R}\hat{b}_{R} = N_{B}/2$ (see Appendix \ref{Appendix_parity}). This leads to the invariant nature of $\rho^{B}_{1}(x)$ against variations of the interaction strengths. Importantly, we note that such a property of $\rho^{B}_{1}(x)$ holds for all different types of residue decay. Apart from that, we also notice that both $|C^{B}_{1,k}|^{2}$ and $|C^{B}_{2,k}|^{2}$ follow a binomial distribution with their maximal values located at $k = 50$ and $k = 49$, respectively [cf. Fig.~\ref{Sorb_BH_NS_basis_type_I} (a) red solid line and blue dashed line]. We note that, this result directly stems from the above-described interspecies two-body phase separation [see also Fig.~\ref{dmats_type_I} (e)]. 

In Appendix \ref{Appendix_A}, we show that $|\Psi_{T-I}\rangle$ [see also the results from Eq.~\eqref{psi_type_I} to Eq.~\eqref{NO_B_type_I}] is exactly equivalent to
\begin{equation}
	|\Psi_{T-I} \rangle = \frac{1}{\sqrt{2 N_{B}!}} \left[\hat{a}_{L}^{\dagger}(\hat{b}_{R}^{\dagger})^{N_{B}} + \hat{a}_{R}^{\dagger}(\hat{b}_{L}^{\dagger})^{N_{B}} \right] |0\rangle, \label{Psi_II_LR}
\end{equation}
with $\hat{a}_{L,R} = \frac{1}{\sqrt{2}} (\hat{a}_{0}^{I} \pm \hat{a}_{1}^{I})$ and $\hat{a}_{i}^{I}$ ($i=0,1$) referring to the annihilation operator acting on the $i$th single-particle state $|\phi_{i}^{I}\rangle$ of the impurity. Remarkably, Eq.~\eqref{Psi_II_LR} essentially represents a Schr\"odinger-cat state being a superposition of two macroscopic many-body states. Each of them corresponds to a configuration where the impurity resides in one well while all the bosons are located in another well. Thus it reflects the above-described two-body phase separation between the two species. We remark that such a Schr\"odinger-cat state has been reported in Ref.~\cite{Impurity_BH_2} by restricting the impurity in the lowest band of the DW (see below). It has also been argued that this state is extremely sensitive to the environment meaning that an arbitrary small perturbation can lead to a collapse onto one of the two macroscopic configurations. This should be indeed the case upon the experimental measurement of this configuration which is a manifestation of the entanglement. The obtained asymptotic GS wave function $|\Psi_{T-I}\rangle$ allows for further explicating the OC of the polaron for strong impurity-bath repulsion: since for the impurity $\langle \psi_{2}^{I} |\phi_{0}^{I}\rangle = \langle \phi_{1}^{I} |\phi_{0}^{I} \rangle = 0$ [cf. Eq.~\eqref{NO_I_type_I}], $|\psi_{2}^{I}\rangle |\psi_{2}^{B}\rangle$ is trivially orthogonal to the ground state $|\Psi_{0} \rangle = |\phi_{0}^{I}\rangle|\phi_{0}^{B}\rangle$ for $g_{IB}=0$. Moreover, the binomial distribution of the coefficients $|C^{B}_{1,k}|^{2}$ together with the normalization condition $\sum_{k} |C^{B}_{1,k}|^{2} = 1$ render \footnote{The maximal value of the binomial distribution of $|C^{B}_{1,k}|^{2}$ is located at $k = N_{B}/2$. For $N_{B} \rightarrow \infty$, the $C^{B}_{1,k}$ value for both $k\ll N_{B}/2$ and $k\gg N_{B}/2$ approaches $C^{B}_{1,k} \to 0$. This can also been seen in Fig.~\ref{Sorb_BH_NS_basis_type_I} (a).} $C^{B}_{1,0}\rightarrow 0$ for $N_{B} \rightarrow \infty$.
As a result, we have $\langle\psi_{1}^{B}|\phi_{0}^{B}\rangle = 0$ [cf. Eq.~\eqref{SO_expansion}] and $|\psi_{1}^{I}\rangle |\psi_{1}^{B}\rangle$ is orthogonal to $|\Psi_{0} \rangle$ as well. In this way, we explicate the above OC of the polaron.

Before closing this section, let us elaborate on the polaron decay for $0<g_{IB}<0.8$ in more detail. As we have already mentioned above, for all values of $g_{IB}$ only the first two Schmidt orbitals are populated in the GS [see Fig.~\ref{EE_band_Impurity_type_I} (a) as well as its inset providing a magnification within the region $0<g_{IB}<0.01$]. The GS wave function correspondingly takes the form  
\begin{equation}
	|\Psi \rangle = \sqrt{\lambda_{1}} |\psi_{1}^{I}\rangle |\psi_{1}^{B}\rangle +  \sqrt{\lambda_{2}} |\psi_{2}^{I}\rangle |\psi_{2}^{B}\rangle.\label{psi_type_I_small_g}
\end{equation}
For the impurity, we have also verified that the Schmidt orbitals follow the relations in Eq.~\eqref{NO_I_type_I} exactly for varying $g_{IB}$. This observation, on the one hand, reveals that the impurity is predominately restricted within the lowest band of the DW which, in turn, explicates the invariant profile of $\rho^{I}_{1}(x)$ for different $g_{IB}$ [cf. Fig.~\ref{dmats_type_I} (a), (b) and the accompanied discussion above]. On the other hand, it also results in an even (odd) parity bosonic Schmidt orbital $|\psi_{1}^{B}\rangle$ ($|\psi_{2}^{B}\rangle$) [see the expansion coefficients $|C^{B}_{1,k}|^{2}$ and $|C^{B}_{2,k}|^{2}$ in Fig.~\ref{Sorb_BH_NS_basis_type_I} (b), (c)]. Moreover, we note that the GS of the mixture deforms from $|\Psi_{0} \rangle$ with respect to $g_{IB}$ as follows: For $0<g_{IB}<0.003$, a larger $g_{IB}$ leads to the intraband excitations of the impurity within the lowest-band of the DW identified by an SP transition of the impurity between the states $|\phi_{0}^{I}\rangle$ and $|\phi_{1}^{I}\rangle$ [see Fig.~\ref{EE_band_Impurity_type_I} (b)]. 
This process is accompanied by a rapid growth of the bath species populations onto the low-lying excited states $\{|\phi_{k}^{B}\rangle \}$ of $\hat{H}_{B}$ for $k \leqslant 12$ [cf. Fig.~\ref{Sorb_BH_NS_basis_type_I} (c)]. 
Consequently, it leads to the sharp decrease of the polaronic residue as depicted in Fig.~\ref{residual_Z} (d). Further increasing the impurity-bath repulsion to $g_{IB}=0.8$, the impurity ceases to be excited to higher bands. Instead, the bosonic species gradually populate its higher-lying excited states $\{|\phi_{k\gg1}^{B}\rangle \}$ [cf. Fig.~\ref{Sorb_BH_NS_basis_type_I} (b)].  
Notice, however, that the bosons are always restricted to the lowest-band of the DW due to the employed two-mode approximation [cf. Eq.~\eqref{2_mode_psi}]. 
The excited states are $\{|\phi_{k}^{B}\rangle \}$ strictly referred to as the eigenstates of the BH Hamiltonian $\hat{H}_{B}$. 
Owing to the fact that the impurity always remains in the lowest band of the DW (when $g_{BB}\approx 0$) during the whole process of the polaron decay, hereafter, we refer to this regime as the intraband excitation induced polaron decay.
 
\subsection{Type-II polaron decay} \label{Type-II_residue_decay} 
\begin{figure*}
	\centering
	\includegraphics[width=1.0\textwidth]{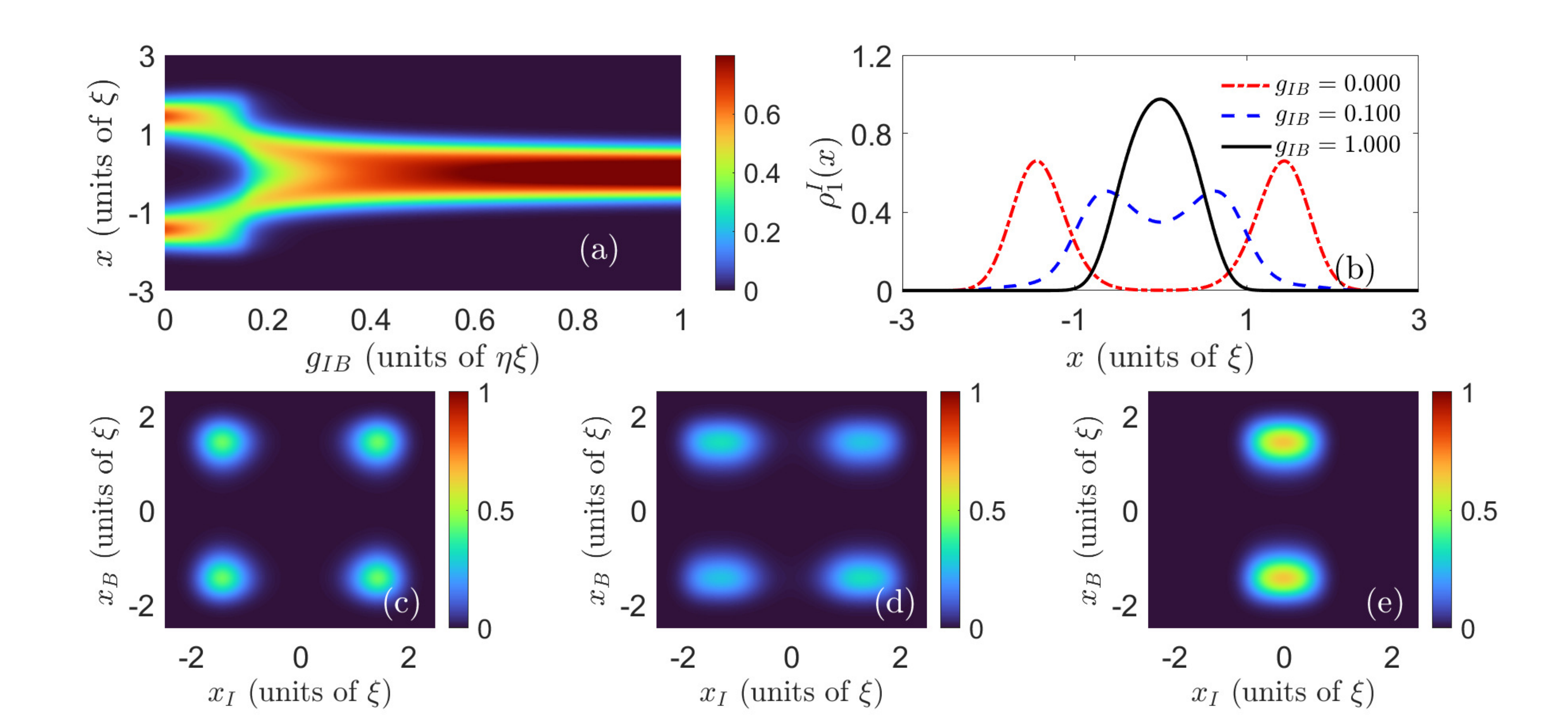}\hfill
	\caption{(Color online) The impurity's reduced densities regarding the type-II residue decay. (a) The single-particle density of the impurity as a function of $g_{IB}$ for $g_{BB} = 0.008$. (b) Profiles of $\rho^{I}_{1}(x)$ for $g_{BB} = 0.008$ and $g_{IB} = 0$ (red dash-dotted line), $g_{IB} = 0.2$ (blue dashed line) and $g_{IB} = 1.0$ (black solid line). The impurity localizes around the central region of the DW for stronger $g_{IB}$ experiencing a one-body phase separation with the medium. (c-e) The corresponding two-body impurity-medium density spatial distributions for the parameters showcased in (b).}
	\label{dmats_type_II}
\end{figure*}

Next, we turn into the description of the type-II polaron decay, which occurs in the interaction regime $g_{BB} \in [g_{BB}^{cI}, g_{BB}^{cII}]$. For $\beta = 1$, we have  $g_{BB}^{cI} = 0.0035$ and $g_{BB}^{cII} = 0.035$. As a characteristic example, we present the density distributions of the impurity for fixed $g_{BB} = 0.008$ and different values of $g_{IB}$ [cf. Figs.~\ref{dmats_type_II} (a) and \ref{dmats_type_II} (b)]. Before proceeding, let us remark that the corresponding bosonic density $\rho^{B}_{1}(x)$ has the same shape as the one depicted in Fig.~\ref{dmats_type_I} (b), i.e. it distributes among the individual wells. On the contrary, the spatial profile of the impurity's density in this regime of $g_{BB}$ depends strongly on $g_{IB}$ [cf. Figs.~\ref{dmats_type_I} (a) and \ref{dmats_type_II} (a)]. Instead of maintaining the initial two-hump structure, for increasing $g_{IB}$, the bosons considerably expel the impurity into the center of the DW rendering a highly localized $\rho^{I}_{1}(x)$ distribution around the region $x \in [-1,1]$ for $g_{IB} > 0.1$. This correspondingly leads to a negligible spatial overlap between $\rho^{I}_{1}(x) $ and $\rho^{B}_{1}(x) $ for strong impurity-bath repulsions [see red dash-dotted and black solid lines in Fig.~\ref{dmats_type_II} (b) referring to the cases $g_{IB} = 0$ and $g_{IB} = 1.0$. Moreover, note that the profile of $\rho^{I}_{1}(x)$ for $g_{IB} = 0$ is same with the one of $\rho^{B}_{1}(x)$]. The above process illustrates another type of phase separation, which is clearly visible on the single-particle densities. It is for this reason that below we shall refer to it as the one-body phase separation~\cite{polaron_eff_mass_Mist_Vol}. Complementary, this phase separation is also captured by the corresponding two-body interspecies density since the latter contains all the information of the single-particle density. For increasing $g_{IB}$, we observe that $\rho^{IB}_{2}(x_{I}, x_{B})$ quickly evolves from a checkerboard pattern with both the diagonal and the off-diagonal elements occupied for weak $g_{IB}$ [cf. Fig.~\ref{dmats_type_II} (c), (d)] towards a configuration with dominant populations around the spatial regions $x_{I} \in [-1,1]$ and $x_{B} \approx \pm 1.5$ for $g_{IB} = 1.0$ [cf. Fig.~\ref{dmats_type_II} (e)]. This reflects a configuration where the impurity lies at the center of the DW, while the bosons are located in the outer two wells. In this regard, it manifests the above-mentioned one-body phase separation.

\begin{figure}
	\centering
	\includegraphics[width=0.5\textwidth]{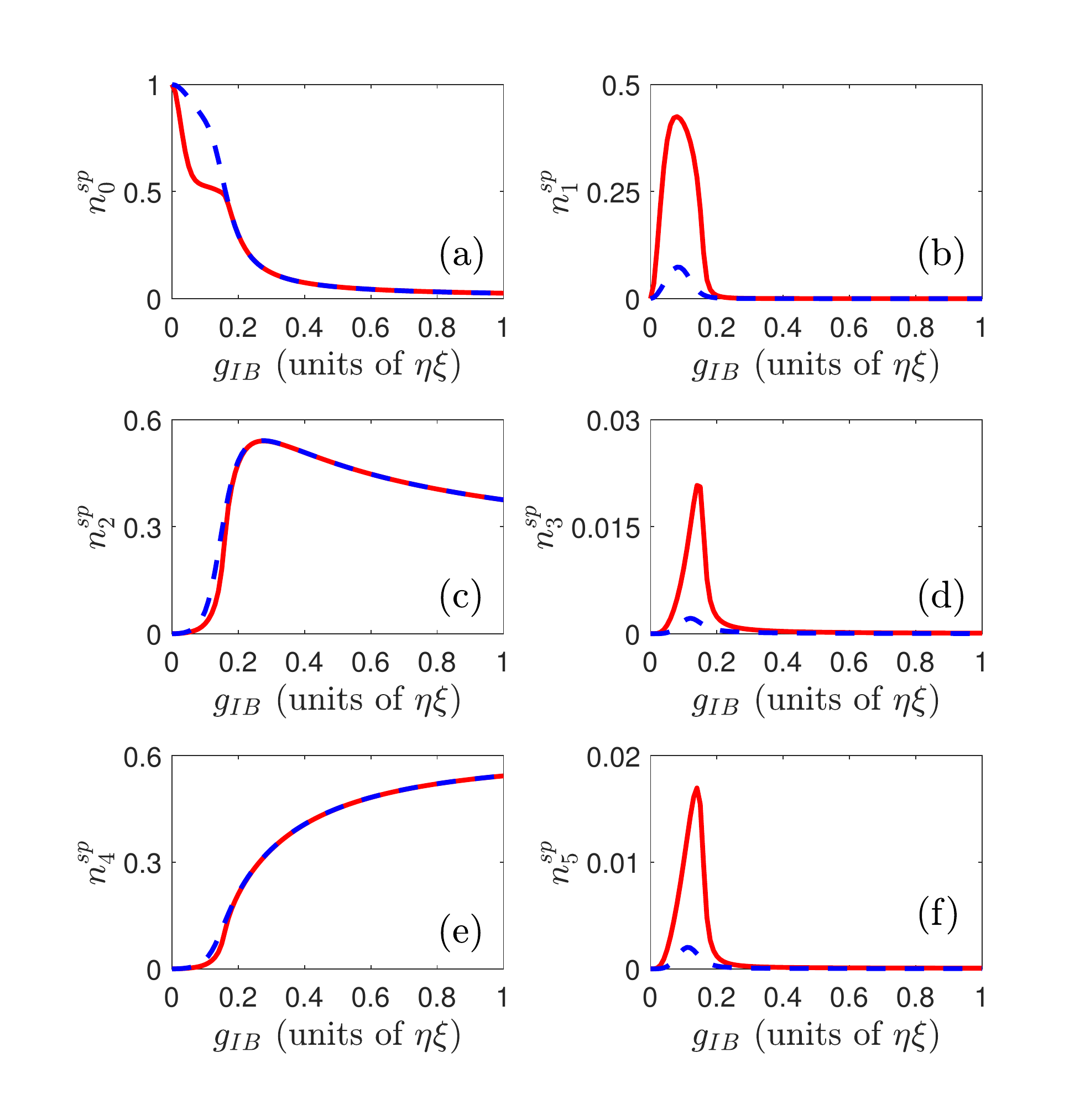}\hfill
	\caption{(Color online) SP occupations of the impurity for the type-II and type-III residue decay regions. The SP occupation $n_{i}^{sp}$ of the impurity with respect to $g_{IB}$ for $g_{BB} = 0.008$ (red solid line) and $g_{BB} = 0.05$ (blue dashed line). The left (right) panels correspond to all even (odd) parity single-particle states.}
	\label{EE_band_Impurity_type_II_III}
\end{figure}

\begin{figure*}
	\centering
	\includegraphics[width=0.9\textwidth]{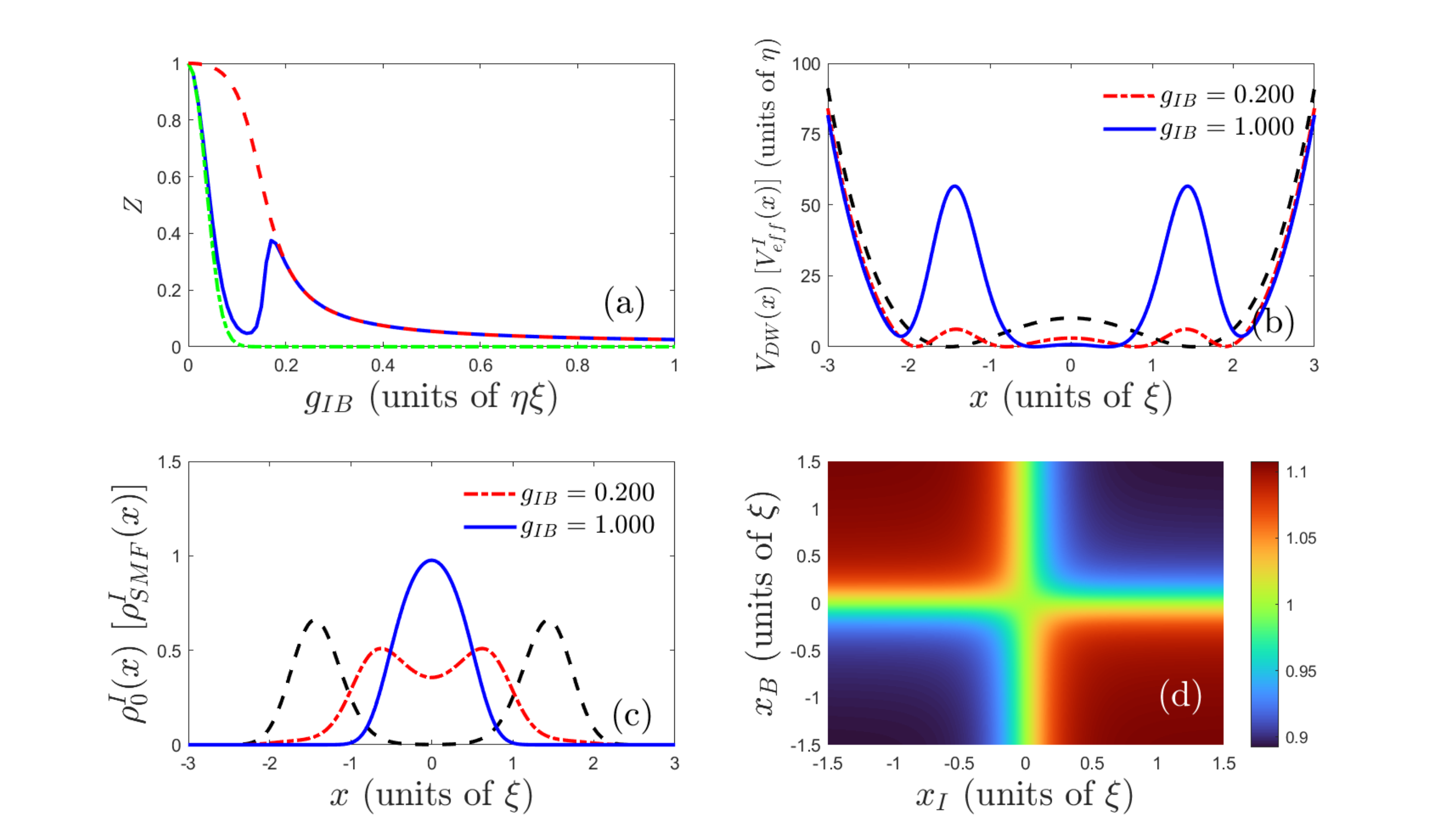}\hfill
	\caption{(Color online) (a) The polaronic residue $Z$ as a function of $g_{IB}$ as predicted within ED for fixed $g_{BB} = 0.008$ (blue solid line). The green dash-dotted line represents the resulting residue $Z$ by restricting the impurity within the lowest band of the DW. The red dashed line denotes the overlap between $|\psi^{I}\rangle$ and $|\phi_{0}^{I}\rangle$ with respect to $g_{IB}$ for $g_{BB} = 0.008$ thus using the SMF method (no entanglement). It can be concluded that the residue predictions of the lowest-band (SMF) are in good agreement with ED for weak (strong) impurity-medium couplings but both approaches fail to predict the correct residue behavior in the intermediate interaction regime.  (b) The effective potential of the impurity for $g_{BB} = 0.008$ and $g_{IB} = 0.2$ (red dash-dotted line) and $g_{IB} = 1.0$ (blue solid line). The black dashed line represents the DW. (c) The corresponding impurity density distributions obtained from the Schmidt orbital $\rho^{I}_{SMF}(x) = |\psi^{I}(x)|^{2}$ (red dash-dotted and blue solid line) for the cases examined in (b). The black dashed line shows the impurity density $\rho^{I}_{0}(x) = |\phi_{0}^{I}(x)|^{2}$ for $g_{IB} = 0$ (see main text). (d) The pair-correlation function for $g_{BB} = 0.008$ and $g_{IB} = 0.1$ obtained from ED simulations, explicating the emergent two-body impurity-medium correlations.}
	\label{eff_Z_Pot}
\end{figure*}

Interestingly, the distinct density behavior of $\rho^{I}_{1}(x) $ in this regime is inherently related to a richer excitation behavior of the impurity as compared to the former region of the type-I polaron decay. In contrast to the case discussed in Sec.~\ref{Type-I_residue_decay}, where the impurity features only the intraband excitations here a larger $g_{IB}$ triggers also interband processes. To demonstrate the latter, we resort again to the above introduced SP occupation $\hat{n}_{i}^{sp}$, which is alternatively related to the underlying band occupation for the impurity as $\hat{n}_{1}^{b} = \hat{n}_{0}^{sp} + \hat{n}_{1}^{sp}$, $\hat{n}_{2}^{b} = \hat{n}_{2}^{sp} + \hat{n}_{3}^{sp}$, and so on. For $g_{IB}<0.1$, the impurity predominately populates the lowest band of the DW, leaving a band occupation $\hat{n}_{1}^{b} > 0.98$ [see Figs.~\ref{EE_band_Impurity_type_II_III} (a) and \ref{EE_band_Impurity_type_II_III} (b) red solid lines]. This is also manifested by restricting the impurity into the lowest band, i.e., by considering the ansatz $\hat{\psi}_{I}(x) = \phi_{0}^{I}(x) \hat{a}_{0}^{I} + \phi_{1}^{I}(x) \hat{a}_{1}^{I}$ and calculate the corresponding polaronic residue $Z$ for different $g_{IB}$. As compared to the numerically exact results, we find a good agreement between each other [cf. Fig.~\ref{eff_Z_Pot} (a), compare blue solid and green dash-dotted lines]. Importantly, we remark that this observation further implies that the corresponding GS wave function $|\Psi\rangle$ takes the form introduced in Eq.~\eqref{psi_type_I_small_g}. In fact, since each Schmidt orbital of the impurity becomes a linear superposition of the SP states $|\phi_{0}^{I} \rangle$ and $|\phi_{1}^{I} \rangle$, to preserve its orthogonality, i.e., $\langle \psi_{i}^{I}|\psi_{j}^{I}\rangle = 0$ for $i \neq j$, there are at most two linearly independent Schmidt orbitals. On the contrary, increasing the impurity-bath repulsion towards $g_{IB} = 0.2$ leads to a sharp decrease (increase) of the SP occupation $\hat{n}_{i}^{sp}$ for $i = 1$ ($i=2,4$) as well as a slow decrease of $\hat{n}_{0}^{sp}$ [cf. Figs.~\ref{EE_band_Impurity_type_II_III} (b), (c) and (e), red solid lines]. This behavior essentially reveals that the impurity experiences transitions among the SP states $|\phi_{1}^{I}\rangle \rightarrow |\phi_{2}^{I}\rangle $ and $|\phi_{1}^{I}\rangle \rightarrow |\phi_{4}^{I}\rangle$. 
Hence, interband excitation processes of the impurity become relevant. Interestingly, combining our observations for the impurity's SP occupations [cf. Fig.~\ref{EE_band_Impurity_type_II_III}] and its density [cf. Fig.~\ref{dmats_type_II} (a)] we conclude that the triggered interband excitations are accompanied by the occurrence of one-body phase separation between the two species. 

Further increasing the impurity-bath interaction to $g_{IB} > 0.2$, a negligible (large) SP occupation of the odd (even) parity states of the impurity is observed. 
Particularly, a slow decrease (increase) of the population $\hat{n}_{0}^{sp}$ and $\hat{n}_{2}^{sp}$ ($\hat{n}_{4}^{sp}$) takes place.
The interband excitations of the impurity are consequently dominated by the transitions e.g. among the SP states $|\phi_{0}^{I}\rangle \leftrightarrow |\phi_{2}^{I}\rangle $, $|\phi_{0}^{I}\rangle \leftrightarrow |\phi_{4}^{I}\rangle$ and $|\phi_{2}^{I}\rangle \leftrightarrow |\phi_{4}^{I}\rangle $, which preserve the parity symmetry of the impurity's wave function (see also the discussion below). 
Importantly,  the Schmidt numbers for $g_{IB}>0.2$ become as $\lambda_{1} \approx 1$ and $\lambda_{i>1} \approx 0$. 
Hence, the GS wave function $|\Psi\rangle$ possesses a simple product form 
\begin{equation}
	|\Psi \rangle = |\psi^{I}\rangle |\psi^{B}\rangle. \label{psi_SMF}
\end{equation}
Therefore, our mixture can be fully captured within the SMF description where the mutual impact of the species is merely an effective potential \cite{polaron_ind_int_3}. The SMF description reduces the interspecies correlated setting into a single-species problem whose effective Hamiltonian reads
\begin{equation}
	\hat{H}_{\text{eff}}^{\sigma} = \hat{H}_{\sigma} + \langle \psi^{\bar{\sigma}}| \hat{H}_{IB}| \psi^{\bar{\sigma}}\rangle,
\end{equation}
with $\bar{\sigma} = B(I)$ for $\sigma = I(B)$. Moreover, the GS of $\hat{H}_{\text{eff}}^{\sigma}$ is equivalent to the Schmidt orbital $|\psi^{\sigma}\rangle$ given by Eq.~\eqref{psi_SMF}.  In this way, the $\sigma$-species particle experiences, instead of the initial DW, an effective potential due to the presence of the other species. For the impurity, this effective potential takes the form
\begin{equation}
V_{\text{eff}}^{I}(x) = V_{DW}(x)  + g_{IB} ~ N_{B}\rho_{1}^{B}(x), \label{V_eff_I}
\end{equation}
which is simply the DW superimposed to a potential proportional to the bosonic density. With increasing $g_{IB}$, we observe that $V_{\text{eff}}^{I}(x)$ deforms from the initial DW towards a triple-well structure for $g_{IB} = 1.0$ with two dominant barriers located at the positions $x\approx \pm 1.5$ [cf. Fig.~\ref{eff_Z_Pot} (b)]. 
The central barrier of the DW is gradually smeared out in the resulting effective potential $V_{\text{eff}}^{I}(x)$ due to the growth of the potential term $g_{IB} ~ N_{B}\rho_{1}^{B}(x)$ for larger $g_{IB}$ \footnote{Note that the corresponding zero-point energies are excluded from the depicted effective potentials in Fig.~\ref{eff_Z_Pot} (b) and thus $V_{\text{eff}}^{I}(x)$ have the same minimum independently of $g_{IB}$.}.  As a result, the impurity becomes significantly localized at the center of the DW [cf. Fig.~\ref{dmats_type_II} (b), black solid line]. Before proceeding, we note that within the parameter space that we have examined, the effective potential for the bosonic species only slightly deviates from the initial DW due to the large particle number imbalance. Hence, the corresponding bosonic Schmidt orbital $|\psi^{B}\rangle$ resembles the GS of the Hamiltonian $\hat{H}_{B}$, i.e.,
\begin{equation}
|\psi^{B}\rangle \approx |\phi_{0}^{B}\rangle. \label{psi_SMF_B}
\end{equation} 
Note, however, that the case $g_{IB} \rightarrow \infty$ will result in a bosonic effective potential significantly different from the DW. This could subsequently render the employed two-mode approximation insufficient for describing the bosonic gas [cf. Eq.~\eqref{2_mode_psi}]. A discussion of this limit, is therefore, beyond the scope of our study. 

The SMF description not only introduces a significant simplification in the study of mixtures but also allows for profound insights concerning the polaron decay as well as its OC. Indeed, the results provided in Eq.~\eqref{psi_SMF} and Eq.~\eqref{psi_SMF_B} suggest that the residue $Z$ for $g_{IB}>0.2$ simply becomes the overlap between the Schmidt orbital $|\psi^{I}\rangle$, i.e., the GS of the effective potential $V_{\text{eff}}^{I}(x)$, and the initial SP state $|\phi_{0}^{I}\rangle$ [cf. Fig.~\ref{eff_Z_Pot} (a) blue solid and red dashed lines].  Upon the increase of $g_{IB}$, the deformation of the density profile $\rho^{I}_{SMF}(x) = |\psi^{I}(x)|^{2}$ from $\rho^{I}_{0}(x) = |\phi_{0}^{I}(x)|^{2}$ [cf. Fig.~\ref{eff_Z_Pot} (c)] suggests the subsequent decrease of the overlap between $|\psi^{I}\rangle$ and $|\phi_{0}^{I}\rangle$, and hence, leads to the corresponding residue decay [see Eq.~(\ref{SO_expansion}) and Fig.~\ref{eff_Z_Pot} (a)]. Based on the behavior of the SP occupations presented in Fig.~\ref{EE_band_Impurity_type_II_III}, we know that the impurity is dominated by interband transitions during this process. In this regard, we refer to this decay mechanism as the interband excitation induced polaron decay.

Turning to the case of $g_{IB} = 1.0$, which corresponds to the OC of the polaron we note that the negligible spatial overlap between $\rho^{I}_{SMF}(x)$ and $\rho^{I}_{0}(x)$ [cf. Fig.~\ref{eff_Z_Pot} (c), blue solid line and black dashed line] indicates that $\langle\phi_{0}^{I}|\psi^{I}\rangle \approx 0$. Together with the fact that $\rho^{I}_{0}(x) = \rho^{B}_{1}(x)$, the above result implies that $\rho^{I}_{SMF}(x)\rho^{B}_{0}(x) \approx 0$, which reflects the above-mentioned one-body interspecies phase separation [cf. Figs.~\ref{dmats_type_II} (a), (b) and (e)]. 
In this way, we obtain the physical connection between the OC of the polaron and the one-body phase separation. With this knowledge, let us point out the asymptotic form of the GS wave function corresponding to this strongly interacting regime
\begin{equation}
	|\Psi_{T-II}\rangle = |\psi^{I}\rangle |\psi^{B}\rangle~~~\text{and} ~~~\rho^{I}_{1}(x)\rho^{B}_{1}(x) = 0, \label{psi_type_II}
\end{equation} 
which accounts for the above OC of the polaron. Interestingly, for a fixed large $g_{IB}$ (e.g., $g_{IB} = 1.0$) and for increasing $g_{BB}$, a transition between the GS wave function $|\Psi\rangle = | \Psi_{T-I}\rangle$ for $g_{BB}< g_{BB}^{cI}$ and $|\Psi\rangle = | \Psi_{T-II}\rangle$ for $g_{BB}> g_{BB}^{cI}$ is realized. This crossover is also accompanied by the deformation of the impurity's density profile from the $\rho^{I}_{1}(x)$ in Fig.~\ref{dmats_type_I} (b) (blue dashed line) to the one depicted in Fig.~\ref{dmats_type_II} (b) (black solid line). In Sec.~\ref{two_bosonic_repulsion}, we will elaborate on the physical origin of this crossover via a spectral analysis of the wave functions $|\Psi_{T-I}\rangle$ and $|\Psi_{T-II}\rangle$. Moreover, it is worth to be mentioned that in the limiting case $N_{B} \rightarrow \infty$ the effective potential in Eq.~\eqref{V_eff_I} can be approximated by $V_{\text{eff}}^{I}(x) \approx g_{IB} ~ N_{B}\rho_{1}^{B}(x)$. The impurity thus feels two infinitely high potential barriers located at positions $x\approx \pm 1.5$, which corresponds to the two-hump structure of the bosonic density [cf. Fig.~\ref{dmats_type_I} (b)]. As a result, it leads to a vanishing overlap between $\rho^{I}_{SMF}(x)$ and $\rho^{B}_{0}(x)$ leaving the residue $Z \rightarrow 0$. 
Another important consequence of the SMF description is that, due to the preservation of the parity symmetry in the effective potential picture, the Schmidt orbital $|\psi^{I}\rangle$ has always even parity. As a result, it leads to the negligible SP occupations $\hat{n}_{i}^{sp}$ of all the odd parity states $|\phi_{i}^{I}\rangle$ for $g_{IB}>0.2$ [see the red solid lines in Fig.~\ref{EE_band_Impurity_type_II_III}  (b), (d), (f)].

\subsection{Type-III polaron decay} \label{Type-III_residue_decay}
Finally we explore the polaron decay in the interaction region $g_{BB}>g_{BB}^{cII}$. For a fixed $g_{IB}>0.2$, the negligible discrepancies between the residue values for the case $g_{BB} = 0.008$ and $g_{BB} = 0.05$ suggest that the GS wave function takes the same product form as introduced in Eq.~\eqref{psi_SMF} [cf. Fig.~\ref{residual_Z} (d) blue solid and black solid line]. This is indeed confirmed by inspecting the $g_{IB}$-dependence of the Schmidt numbers finding that $\{\lambda_{i \geqslant2}\}$ have negligible values for $g_{BB} = 0.05$ and $g_{IB} > 0.2$ (results not shown). It turns out that the above relation holds for all cases with $g_{IB}< 0.2$ as well. As a matter of fact, the mixture in the region $g_{BB}>g_{BB}^{cII}$ is fully described by the SMF ansatz. The corresponding polaron decay with increasing $g_{IB}$ can thus be readily interpreted via the above introduced effective potential. As expected, it gives rise to the same asymptotic form of the GS wave function $|\Psi_{T-II}\rangle$ as introduced in Eq.~\eqref{psi_type_II} for large $g_{IB}$ as well as the negligible SP occupations $\hat{n}_{i}^{sp}$ for all odd parity states $|\phi_{i}^{I}\rangle$ [see the blue dashed lines in Fig.~\ref{EE_band_Impurity_type_II_III} (b), (d) and (f)]. Due to the small variations of the effective potential for bosons with respect to the initial DW, we note that for the bosonic species the relation in Eq.~\eqref{psi_SMF_B} holds as well, indicating the negligible probability for the bosons to populate the excited states $\{|\phi_{i>0}^{B}\rangle\}$.

Albeit the fact that, for a fixed $g_{IB}$, there are no significant differences among the reduced densities $\rho^{\sigma}_{1}(x)$ and $\rho^{IB}_{2}(x_{I}, x_{B})$ between the case of $g_{BB}>g_{BB}^{cII}$ and the one for $g_{BB} \in [g_{BB}^{cI}, g_{BB}^{cII}]$, prominent deviations are imprinted in the corresponding pair-correlation function. 
The latter is defined as
\begin{equation}
g_{2}(x_{I}, x_{B}) = \frac{\rho^{IB}_{2}(x_{I}, x_{B})}{\rho^{I}_{1}(x_{I})\rho^{B}_{1}(x_{B})}.
\end{equation}
Notice that, through the division by the single-particle densities the $g_2$ function, as compared to the two-body density $\rho^{IB}_{2}(x_{I}, x_{B})$, naturally excludes the impact of the spatial inhomogeneity. It can be also readily deduced that within the SMF description $g_{2}$ is simply unity due to the product form of the many-body wave function. In contrast, for $g_{IB} = 0.1$ and $g_{BB} = 0.008$ the pair-correlation function deviates significantly from unity with $g_{2}<1$ ($g_{2}>1$) along the diagonal (off-diagonal) region. This pattern signifies the presence of strong anti-correlations between the impurity and the majority bosons [cf. Fig.~\ref{eff_Z_Pot} (d)].

\subsection{Origin of the critical bosonic repulsion } \label{two_bosonic_repulsion}

So far, we have discussed the properties of the three different types of residue decay in terms of $g_{BB}$ upon increasing the repulsive impurity-bath interaction strength. It has been found that the corresponding residue regimes are characterized by two critical bosonic repulsion strengths $g_{BB}^{cI}$ and $g_{BB}^{cII}$. In addition, two asymptotic GS wave function forms namely $| \Psi_{T-I}\rangle$ and $| \Psi_{T-II}\rangle$ were obtained for strong impurity-bath repulsion, which in turn account for the OC of the polaron. Interestingly, for a fixed large $g_{IB}$ (e.g., $g_{IB} = 1.0$) and increasing $g_{BB}$, we observe  an abrupt transition of the GS wave function from $|\Psi\rangle = | \Psi_{T-I}\rangle$ for $g_{BB}< g_{BB}^{cI}$ towards $|\Psi\rangle = | \Psi_{T-II}\rangle$ for $g_{BB}> g_{BB}^{cI}$. Hereafter, we will first perform a spectral analysis with respect to the above two asymptotic wave functions. This allows us to unveil the origin of the critical bosonic repulsion $g_{BB}^{cI}$. Afterwards, we comment on the existence of $g_{BB}^{cII}$ and demonstrate how the impurity-medium mass ratio affects their values.

To begin with, we note that for both $| \Psi_{T-I}\rangle$ and $| \Psi_{T-II}\rangle$ the corresponding interspecies interaction energies given by $\langle \Psi_{T-I}|\hat{H}_{IB}| \Psi_{T-I}\rangle$ and $\langle \Psi_{T-II}|\hat{H}_{IB}| \Psi_{T-II}\rangle$ can be safely neglected. In fact, due to the phase separation occurring on either the one-body or the two-body level in the limit $g_{IB}\rightarrow 1$, atoms from different species have negligible probability to reside in the vicinity of one another. Together with the employed contact form of the interspecies interaction, it thereby results in a negligible interspecies interaction energy. 
In this regard, we solely need to compute the energies associated with the single-species Hamiltonians $\hat{H}_I$ and $\hat{H}_B$ [Eq.~(\ref{Hamiltonian_IB})]. For $| \Psi_{T-I}\rangle$, we have 
\begin{equation}
	E_{T-I} = \langle\Psi_{T-I}| \hat{H}_{I} | \Psi_{T-I}\rangle + \langle\Psi_{T-I}| \hat{H}_{B} | \Psi_{T-I}\rangle = E_{SW}^{I} + E_{SW}^{B}, \label{E_T_I}
\end{equation}
with $E_{SW}^{I}$ ($E_{SW}^{B}$) being the energy corresponding to the configuration where the impurity (all the bosons with a fixed $g_{BB}$) are restricted in either the left (right) or the right (left) well.
On the other hand, for $| \Psi_{T-II}\rangle$ the resulting energy reads
\begin{equation}
	E_{T-II} = \langle\Psi_{T-II}| \hat{H}_{I}  | \Psi_{T-II}\rangle + \langle\Psi_{T-II}|  \hat{H}_{B} | \Psi_{T-II}\rangle = E_{SMF}^{I} + E_{0}^{B}, \label{E_T_II}
\end{equation}
in which $E_{SMF}^{I} = \langle \psi^{I} |\hat{H}_{I} |\psi^{I} \rangle$ represents the impurity energy obtained from the SMF Schmidt orbital $|\psi^{I} \rangle$ corresponding to a strong impurity-bath repulsion and fixed $g_{BB}> g_{BB}^{cI}$. Here, we set $g_{IB} = 1.0$ and $g_{BB} = 0.008$. 
We note that as long as $g_{BB}> g_{BB}^{cI}$, different values of $g_{BB}$ do not impact the shape of the impurity's effective potential, thus resulting in the same Schmidt orbital $|\psi^{I} \rangle$ (see also the discussion in Sec.~\ref{Type-II_residue_decay}).  Also, $E_{0}^{B}$ is the GS energy of the Hamiltonian $\hat{H}_{B}$ for a fixed $g_{BB}$. 

\begin{figure}
	\centering
	\includegraphics[width=0.5\textwidth]{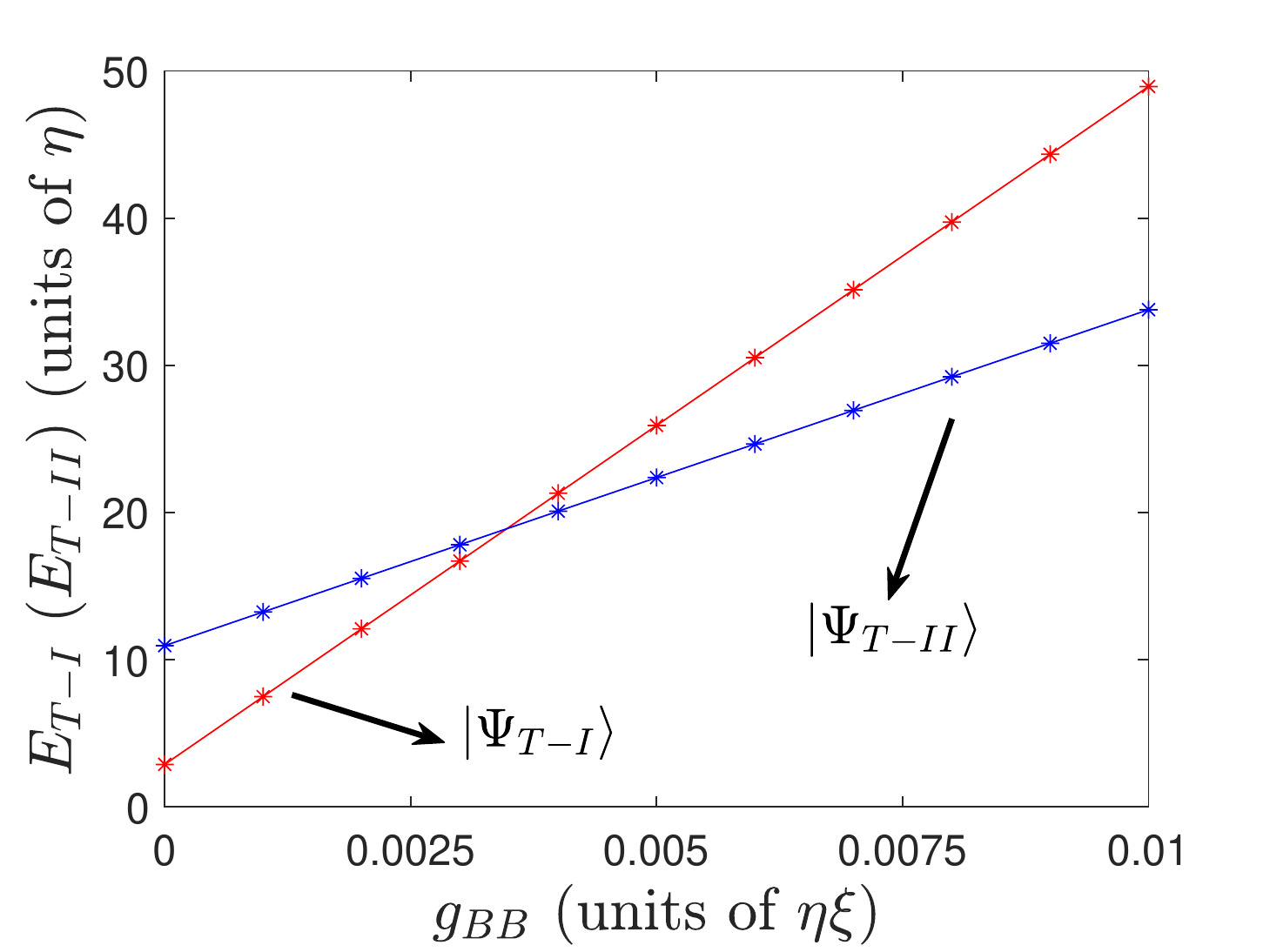}\hfill
	\caption{(Color online) Energy competition between two asymptotic GS wave functions (see main text). The energies $E_{T-I}$ (red solid line) and $E_{T-II}$ (blue solid line) obtained within ED as a function of $g_{BB}$. Here, $E_{T-I} = \langle\Psi_{T-I}| \hat{H}_{I} + \hat{H}_{B} | \Psi_{T-I}\rangle $ and $E_{T-II} = \langle\Psi_{T-II}| \hat{H}_{I} + \hat{H}_{B} | \Psi_{T-II}\rangle$. }
	\label{E_effective_WFs}
\end{figure}

Eqs.~\eqref{E_T_I} and \eqref{E_T_II} essentially reveal that both $E_{T-I} $ and $E_{T-II} $ depend on the bosonic repulsion strength $g_{BB}$. Therefore, the competition between the two energies upon the increase of $g_{BB}$ directly determines the form of the GS wave function $|\Psi\rangle$ in the asymptotic limit of strong $g_{IB}$. In the case of $g_{BB} \approx 0$, the negligible width of the lowest band of the DW leads to $E_{SW}^{I} \approx E_{0}^{I}$ and $E_{SW}^{B} \approx E_{0}^{B}$ with $E_{0}^{I}$ being the GS energy of $\hat{H}_{I}$. Since $E_{SMF}^{I}$ is obviously larger than $E_{0}^{I}$, it thus results in $E_{T-I}<E_{T-II}$. Thereby, the wave function $|\Psi_{T-I}\rangle$ is energetically more favorable. In contrast, with increasing $g_{BB}$, the rapid growth of the on-site bosonic repulsion renders the energy $E_{SW}^{B}$ dominant as compared to all other energies. As a result, the GS of the mixture turns out to be $|\Psi \rangle = |\Psi_{T-II}\rangle$. Based on this knowledge, we can deduce that the critical bosonic repulsion strength $g_{BB}^{cI}$ corresponds to the case where $E_{T-I} = E_{T-II}$ holds [cf. Fig.~\ref{E_effective_WFs}]. 
Turning to $g_{BB}^{cII}$, it corresponds to the situation that the GS of the mixture always acquires the product form $|\Psi \rangle = |\psi^{I}\rangle |\psi^{B}\rangle$ for different $g_{IB}$, in which $ |\psi^{B}\rangle \approx |\phi_{0}^{B}\rangle $ [cf. Eqs.~\eqref{psi_SMF} and \eqref{psi_SMF_B}]. Thus, the bosonic species is hardly excited irrespectively of the variations of the interspecies coupling. 
Accordingly, it becomes clear that only a large enough $g_{BB}$ can result in corresponding large energy differences between the GS and the excited states of $\hat{H}_{B}$. 
This fact effectively prohibits the excitations among the bosons. 

Finally, let us demonstrate how the variations of mass-imbalance impact the values of the above-discussed bosonic repulsions. For a light impurity, the associated SP wave functions $\{\phi_{i}^{I}(x)\}$ are much more spatially extended as compared to the ones for a heavy impurity (results are not shown). This facilitates the interband excitations of the impurity quantified by the transition amplitude $T^{I} = \sum_{i>1} U_{i}$ with
\begin{equation}
U_{i} = \sum_{j = 0,1} \sum_{k,l = 0,1} g_{IB} \int dx ~ |\phi_{i}^{I}(x) \phi_{j}^{I}(x) \varphi_{k}^{B}(x) \varphi_{l}^{B}(x)|^{2}.
\end{equation}
It represents the magnitude of the transition amplitude of the impurity, due to the impurity-bath coupling, between the energetically lowest and higher bands of the DW after averaging out different configurations of the bosons. 
Recall that, from the discussions in Sec.~\ref{Type-II_residue_decay} and Sec.~\ref{Type-III_residue_decay}, it is known that $g_{BB}^{cII}$ essentially corresponds to the regime where the impurity is dominated by interband excitation processes upon increasing $g_{IB}$. Hence, it gives rise to a smaller $g_{BB}^{cII}$ in the case of $\beta = 0.5$ when compared to $\beta = 2.0$. Regarding $g_{BB}^{cI}$, it essentially characterizes the transition of the GS wave function from $|\Psi_{T-I}\rangle$ to $|\Psi_{T-II}\rangle$ for larger $g_{IB}$. As we have already discussed above, the interspecies interaction energy in this limit always vanishes. 
Accordingly, changes of $T^{I}$ lead to a negligible impact on this GS crossover. 
It is for this reason that $g_{BB}^{cI}$ is inert to variations of the mass-imbalance $\beta$. 

\section{Conclusions and Outlook} \label{Conclusions}
We have investigated the polaronic properties in the ground state of a binary atomic mixture consisting of a single impurity and bosonic gas confined in a 1D DW. Due to the trap-induced spatial imhomogeneity, three distinct polaronic residue regions in terms of the bosonic repulsion have been identified with respect to the impurity-bath interaction strength. They are characterized by two critical values of the bosonic repulsion denoted as $g_{BB}^{cI}$ and 
$g_{BB}^{cII}$. Moreover, the increase of $g_{IB}$ always results in a suppressed  residue, signifying the occurrence of the OC of the polaron. Depending on the value of $g_{BB}$ being below or above $g_{BB}^{cI}$ or $g_{BB}^{cII}$, the residue exhibits a sharp decrease, an initial decrease followed by a pronounced revival or a slow monotonous decay, respectively. The presence of these interaction regions stems from the interplay between the intra- and interband excitations of the impurity and are clearly imprinted in the structure of the corresponding reduced density matrices. The latter essentially reveal a phase separation on either the one- or the two-body level depending on the bosonic repulsion. Additionally, it is found that the interspecies mass ratio affects the values of the $g_{BB}^{cII}$ critical bosonic repulsion while $g_{BB}^{cI}$ appears to be almost un-affected. It is argued that this behavior can be understood in terms of the respective interband transition amplitudes.

To provide deeper insights into the polaron decay and the existence of the interaction regions where the residue exhibits a distinct behavior, we perform a detailed analysis of the structure of the many-body wave function at specific limits. By utilizing a Schmidt decomposition of the many-body wave function, it allows us to construct two asymptotic configurations of the GS wave function for strong interspecies repulsion. Importantly, they also account for the observed OC of the polaron. Finally, by means of a spectral analysis, we elucidate the physical origin of the observed two critical bosonic repulsions and demonstrate how the mass-imbalance impacts their values. 

Possible future research directions include the investigation of the polaronic properties arising in the ground state of the binary atomic mixture in the presence of long-range e.g. dipolar interactions. 
Another interesting perspective is to study the polaron dynamics following an interaction quench. Here, the impact of higher-band excitations of the impurity or a beyond the two-site Bose-Hubbard description for the bosonic species would be worthwhile to pursue.

\begin{acknowledgments}
This work has been funded by the Deutsche Forschungsgemeinschaft (DFG, German Research Foundation) - SFB 925 - project 170620586. S.I.M. gratefully acknowledges financial support 
from the NSF through a grant for ITAMP at Harvard University. 
The authors acknowledge fruitful discussions with G.M. Koutentakis.
\end{acknowledgments}

\appendix
\section{Impact of an even parity GS wave function $|\Psi \rangle$ to the bosonic single-particle density} \label{Appendix_parity}

Let us demonstrate that an even parity GS wave function $|\Psi \rangle$ essentially implies that the bosonic spatial distribution between the two wells obeys the condition $\hat{b}^{\dagger}_{L}\hat{b}_{L}  = \hat{b}^{\dagger}_{R}\hat{b}_{R} = N_{B}/2$. 
Therefore, it leads to an invariant $\rho^{B}_{1}(x)$ profile against variations of the involved interaction strengths. 
To showcase this behavior, we first rewrite the bosonic annihilation operator as $\hat{b}_{L,R} = \frac{1}{\sqrt{2}} (\hat{a}_{0}^{B} \pm \hat{a}_{1}^{B})$, with $\hat{a}_{i}^{B}$ being the annihilation operator acting on the $i$th SP state $|\varphi_{i}^{B}\rangle$ of the DW. As a result, the occupation number of the bosons on the left (right) well becomes 
\begin{equation}
	N_{L,R}^{B} = \langle \Psi|\hat{b}_{L,R}^{\dagger}\hat{b}_{L,R}|\Psi \rangle = \frac{N_{B}}{2} \pm \frac{1}{2} \langle \Psi|(\hat{a}_{0}^{B})^{\dagger} \hat{a}_{1}^{B} + (\hat{a}_{1}^{B})^{\dagger} \hat{a}_{0}^{B}|\Psi \rangle , \label{N_B_LR}
\end{equation}
where we have used the fact that $(\hat{a}_{0}^{B})^{\dagger} \hat{a}_{0}^{B} + (\hat{a}_{1}^{B})^{\dagger} \hat{a}_{1}^{B} = N_{B}$, due to the particle number conservation and the employed two-mode approximation for the bosons. Further employing a Schmidt decomposition on the GS wave function $|\Psi\rangle$, Eq.~\eqref{N_B_LR} reduces to
\begin{equation}
	N_{L,R}^{B} = \frac{N_{B}}{2} \pm \frac{1}{2} \sum_{i=1}^{\infty} \lambda_{i} \left[\langle \psi_{i}^{B} | (\hat{a}_{0}^{B})^{\dagger} \hat{a}_{1}^{B} +  (\hat{a}_{1}^{B})^{\dagger} \hat{a}_{0}^{B} |\psi_{i}^{B}\rangle \right], \label{N_B_LR_schmidt}
\end{equation}
with the last term in Eq.~\eqref{N_B_LR_schmidt} corresponding to a parity-flipping operation acting on the bosonic species. 
Since the GS wave function $|\Psi\rangle$ is of even parity, each product state $|\psi_{i}^{I}\rangle |\psi_{i}^{B}\rangle$ in Eq.~\eqref{psi_type_I} then has an even parity as well, which renders the bosonic Schmidt orbital $|\psi_{i}^{B}\rangle$ possessing a definitive parity. 
As a result, the last term in Eq.~\eqref{N_B_LR_schmidt} vanishes and thus we have $N_{L,R}^{B} = N_{B}/2$.

\section{Expressing the asymptotic wave function $|\Psi_{T-I} \rangle$ in the Wannier-basis representation} \label{Appendix_A}
In this part, we demonstrate that the wave function $|\Psi_{T-I} \rangle$ of Eq.~\eqref{psi_type_I} is equivalent to 
\begin{equation}
	|\Psi_{T-I} \rangle = \frac{1}{\sqrt{2 N_{B}!}} \left[\hat{a}_{L}^{\dagger}(\hat{b}_{R}^{\dagger})^{N_{B}} + \hat{a}_{R}^{\dagger}(\hat{b}_{L}^{\dagger})^{N_{B}} \right] |0\rangle, \label{psi_type_I_LR}
\end{equation}
where
\begin{equation}
	\hat{a}_{L,R} = \frac{1}{\sqrt{2}} (\hat{a}_{0}^{I} \pm \hat{a}_{1}^{I}),~~~~\hat{b}_{L,R} = \frac{1}{\sqrt{2}} (\hat{a}_{0}^{B} \pm \hat{a}_{1}^{B}), \label{operator_LR_NA}
\end{equation}
with $\hat{a}_{i}^{I}$ ($\hat{a}_{i}^{B}$) being the annihilation operator acting on the $i$th SP state $|\phi_{i}^{I}\rangle$ ($|\varphi_{i}^{B}\rangle$) of the impurity and the bosonic species, respectively. By substituting Eq.~\eqref{operator_LR_NA} into Eq.~\eqref{psi_type_I_LR}, it results in $|\Psi_{T-I} \rangle = \frac{1}{\sqrt{2}} \left[|\phi_{0}^{I}\rangle |\tilde{\psi}_{1}^{B}\rangle + |\phi_{1}^{I}\rangle|\tilde{\psi}_{2}^{B}\rangle \right]$, where 
\begin{align}
	|\tilde{\psi}_{1}^{B}\rangle &=\sum_{k=0,2,4,\dots} ^{N_{B}} \tilde{C}^{B}_{1,k} |\phi_{k}^{B}\rangle =  \sum_{k=0,2,4,\dots} ^{N_{B}} (\frac{1}{\sqrt{2}})^{N_{B}-1} \left(\begin{array}{c} N_{B} \\ k \end{array} \right)^{1/2} |\phi_{k}^{B}\rangle, \nonumber \\
	|\tilde{\psi}_{2}^{B}\rangle &=\sum_{k=1,3,5,\dots} ^{N_{B}} \tilde{C}^{B}_{2,k} |\phi_{k}^{B}\rangle =  \sum_{k=1,3,5,\dots} ^{N_{B}} -(\frac{1}{\sqrt{2}})^{N_{B}-1}  \left(\begin{array}{c} N_{B} \\ k \end{array} \right)^{1/2} |\phi_{k}^{B}\rangle. \label{expansion_NO_B_LR}
\end{align}
Here, $|\phi_{k}^{B}\rangle$ denotes the $k$th eigenstate of $\hat{H}_{B}$ for $g_{BB} = 0$. The expansion coefficients $|\tilde{C}^{B}_{1,k}|^{2}$ and  $|\tilde{C}^{B}_{2,k}|^{2}$ in Eq.~\eqref{expansion_NO_B_LR} follow a binomial distribution and match exactly to the $|C^{B}_{1,k}|^{2}$ and $|C^{B}_{2,k}|^{2}$ of Eq.~\eqref{NO_B_type_I} [cf. Fig.~\ref{Sorb_BH_NS_basis_type_I} (a) black solid line, red solid line and blue dashed line]. In this way, we show that $|\Psi_{T-I} \rangle$ in Eq.~\eqref{psi_type_I} is equivalent to the form of Eq.~\eqref{psi_type_I_LR}.
 
\section{The species mean-field description} \label{Appendix_B}
The species mean-field description assumes the wave function of the mixture to have a simple product form, i.e. $|\Psi \rangle = |\psi^{I} \rangle |\psi^{B}\rangle$. This leads to the corresponding Lagrangian of the mixture 
\begin{equation}
	\mathcal{L} = \langle \Psi |\hat{H}|\Psi \rangle +\sum_{\sigma = I,B}  \mu_{\sigma} \left[1-\langle \psi^{\sigma}|\psi^{\sigma}\rangle \right],
\end{equation}
where $\mu_{\sigma}$ is the associated Lagrange  multiplier under the constraint of norm conservation of $|\psi^{\sigma}\rangle$. By utilizing a variational principle with respect to each orbital $|\psi^{\sigma}\rangle $, we immediately obtain a Schr\"odinger-type equation for the $\sigma$-species $\hat{H}^{\sigma}_{\text{eff}} |\psi^{\sigma}\rangle = \mu_{\sigma} |\psi^{\sigma}\rangle$, in which the effective Hamiltonian $\hat{H}^{\sigma}_{\text{eff}}$ reads
\begin{equation}
	\hat{H}^{\sigma}_{\text{eff}} = \hat{H}_{\sigma} + \langle \psi^{\bar{\sigma}}|\hat{H}_{IB} |\psi^{\bar{\sigma}}\rangle, \label{eff_H_SMF}
\end{equation} 
with $\bar{\sigma} = B(I)$ for $\sigma = I(B)$. In this way, we note that the GS of $\hat{H}^{\sigma}_{\text{eff}}$ is equivalent to the Schmidt orbital  $|\psi^{\sigma}\rangle$. Moreover, the last term in Eq.~\eqref{eff_H_SMF} represents a potential 
\begin{equation}
	\hat{V}^{\sigma} = \langle \psi^{\bar{\sigma}}|\hat{H}_{IB} |\psi^{\bar{\sigma}}\rangle = g_{IB} N_{\bar{\sigma}}\int dx ~\rho_{1}^{\bar{\sigma}} (x)  ~ \hat{\psi}^{\dagger}_{\sigma} (x) \hat{\psi}_{\sigma} (x),  \label {induced_potential_SMF}
\end{equation}
with $ \rho_{1}^{\bar{\sigma}} (x) = \langle \psi^{\bar{\sigma}}| \hat{\psi}^{\dagger}_{\bar{\sigma}} (x)\hat{\psi}_{\bar{\sigma}} (x) |\psi^{\bar{\sigma}}\rangle/ N_{\bar{\sigma}}$ being the one-body density of the $\bar{\sigma}$-species. From Eqs.~\eqref{eff_H_SMF} and \eqref{induced_potential_SMF}, we conclude that within the species mean-field description the mutual impact of the species is merely an additional potential experienced by the other species.

\end{document}